\documentclass[12pt]{article}
\usepackage{xcolor} 
\usepackage{tikz} 
\usepackage[utf8]{inputenc}
\usepackage{mathpazo}
\usepackage{placeins}
\usepackage{multirow}
\usepackage[normalem]{ulem}
\usepackage{relsize}
\usepackage{algorithm}
\usepackage[noend]{algpseudocode}
\usepackage{graphicx}
\usepackage{booktabs}
\usepackage{framed}
\usepackage{listings}
\usepackage{float}
\usepackage{lipsum}
\usepackage{mathtools, amsfonts}
\usepackage{siunitx}
\usepackage{amsmath, amssymb} 
\usepackage{bm} 
\usepackage[acronym]{glossaries} 
\usepackage{enumerate}
\usepackage[shortlabels]{enumitem}
\usepackage[colorlinks,allcolors=black,citecolor=blue,urlcolor=blue]{hyperref}
\usepackage[citestyle=apa,style=apa,backend=biber,url=false,uniquename=false, useprefix=true, doi=false]{biblatex}
\AtEveryBibitem{%
  \clearfield{note}%
  \clearfield{urlyear}
  \clearfield{urlmonth}
}
\usepackage{dsfont} 
\DefineBibliographyStrings{english}{%
  circa = {{}ca\adddot},
}

\newlist{steps}{enumerate}{1}
\setlist[steps,1]{label=\textbf{Step \arabic*:}, left=0pt, labelwidth=2em, labelsep=1em, align=left}

\newcommand{\pkg}[1]{{\normalfont\fontseries{b}\selectfont #1}} \let\proglang=\textsf 

\makeatletter
\AtBeginDocument{%
  \expandafter\renewcommand\expandafter\subsection\expandafter
    {\expandafter\@fb@secFB\subsection}%
  \newcommand\@fb@secFB{\FloatBarrier
    \gdef\@fb@afterHHook{\@fb@topbarrier \gdef\@fb@afterHHook{}}}%
  \g@addto@macro\@afterheading{\@fb@afterHHook}%
  \gdef\@fb@afterHHook{}%
}
\makeatother

\title{CLaRe: Compact near-lossless Latent Representations of High-Dimensional Object Data}
\author{
Emma Zohner\thanks{Department of Statistics, Rice University.}
\and
Edward Gunning\thanks{Department of Biostatistics, Epidemiology and Informatics, University of Pennsylvania (Corresponding Author:
\href{mailto:Edward.Gunning@pennmedicine.upenn.edu}{Edward.Gunning@pennmedicine.upenn.edu}).}
\and
Giles Hooker\thanks{Department of Statistics and Data Science, University of Pennsylvania.} 
\and
Jeffrey Morris\thanks{Department of Biostatistics, Epidemiology and Informatics, University of Pennsylvania (Corresponding Author: \href{mailto:Jeffrey.Morris@pennmedicine.upenn.edu}{Jeffrey.Morris@pennmedicine.upenn.edu}).}
}
\date{}

\begin{document}

\maketitle

\begin{abstract}
Latent feature representation methods play an important role in the dimension reduction and statistical modeling of high-dimensional complex data objects.
However, existing approaches to assess the quality of these methods often rely on aggregated statistics that reflect the central tendency of the distribution of information losses, such as average or total loss, which can mask variation across individual observations.
We argue that controlling average performance is insufficient to guarantee that statistical analysis in the latent space reflects the data-generating process and instead advocate for controlling the worst-case generalization error, or a tail quantile of the generalization error distribution.
Our framework, CLaRe (Compact near-lossless Latent Representations), introduces a systematic way to balance compactness of the representation with preservation of information when assessing and selecting among latent feature representation methods. 
To facilitate the application of the CLaRe framework, we have developed GLaRe\footnote{\url{https://github.com/edwardgunning/GLaRe}} (Graphical Analysis of Latent Representations), an open-source \proglang{R} package that implements the framework and provides graphical summaries of the full generalization error distribution.
We demonstrate the utility of CLaRe through three case studies on high-dimensional datasets from diverse fields of application.
We apply the CLaRe framework to select among principal components, wavelets and autoencoder representations for each dataset.
The case studies reveal that the optimal latent feature representation varies depending on dataset characteristics, emphasizing the importance of a flexible evaluation framework.
\end{abstract}

\section{Introduction}

Advancements in computer storage capabilities and computational speed have made high-dimensional complex data ubiquitous in all areas of science.
Efficient analysis of such data often necessitates a lower-dimensional representation in a ``latent" space that retains the salient structure of the original data and is more amenable to statistical modeling.
We use the term \emph{latent feature representation method} to refer to statistical and machine learning approaches that achieve this dimension reduction through a (linear or non-linear) transformation of the data to a lower-dimensional space of features.
Examples of latent feature representation methods include Principal Component Analysis (PCA) \parencite{hotelling_analysis_1933}, wavelet representations \parencite{daubechies_wavelet_1990}, t-distributed stochastic neighbor embedding (t-SNE) \parencite{maaten_visualizing_2008}, uniform manifold approximation and projection (UMAP) \parencite{mcinnes_umap_2020} and autoencoders \parencite{rumelhart_learning_1986}.

These latent representations are routinely used in downstream analysis, e.g., latent features can be employed as predictors in multivariable regression or classification, in clustering, or as the response vector in multivariate regression models \parencite{niu_dimensionality_2011,wang_role_2014, cook_fisher_2007}. 
Therefore, assessing how well these methods preserve information is a pertinent challenge.
Training error (i.e., how well the method reconstructs the training data) provides a naive estimate of information loss that tends to be overly-optimistic.
\emph{Generalization error}, which can be defined as a latent feature representation method's error in reconstructing unseen data (i.e., data not used to learn the representation), is used to more accurately quantify information loss and is typically computed using cross-validation approaches \parencite[see, e.g.,][]{becht_dimensionality_2019, bro_cross-validation_2008, wold_cross-validatory_1978, eastment_cross-validatory_1982,krzanowski_cross-validation_1987, minka_automatic_2000, rajan_bayesian_1994, camacho_cross-validation_2014, diana_cross-validation_2002, hubert_fast_2007, josse_selecting_2012, saccenti_use_2015}.

However, most existing approaches summarize generalization error using a single statistic that is aggregated across all observations (e.g., average or total information loss) and reflects the central tendency of the full distribution of individual losses, which can mask variation across observations.
For example, a satisfactory average performance might hide cases where individual observations are very poorly represented and if a generative statistical model is formulated for the latent features, it might disproportionately favor observations that are reconstructed well.
To ensure that models formulated in the latent space can reflect the true data-generating process, it is important to evaluate the entire distribution of generalization errors and control metrics such as the worst-case performance or quantiles of the generalization error distribution.
In addition, evaluation of a method's information preservation must be balanced with the compactness of the representation, as prioritizing losslessness alone can lead to unnecessarily complex representations, whereas overly compact (``lossy") representations may fail to preserve important dataset characteristics.

In this work, we present \textbf{CLaRe} (compact near-lossless latent representations), a framework designed to assess and select among latent feature representations using the full distribution of generalization errors.
Our approach uses cross-validation to compute this distribution and uses it to evaluate a latent feature representation using a coherent set of user-specified criteria.
The key contributions of our framework are as follows:
\begin{enumerate}
    \item In contrast to conventional approaches focusing on aggregated measures that reflect the central tendency of the error distribution (training or generalization error), the CLaRe framework ensures that a level of error tolerance is met for quantiles of the distribution of generalization error (e.g., worst case or $95$th percentile). Latent feature representation methods are evaluated on their ability to preserve all (or most) of the salient information in a dataset.
    \item The suitability of a latent feature representation depends heavily on the characteristics of the dataset at hand \parencite[Section 3, pp. 325--328]{morris_functional_2015}.
    By defining a consistent set of criteria to evaluate different latent feature representation methods, CLaRe enables objective comparisons among different methods to identify the most suitable latent representation method for a specific dataset and application.
    As we demonstrate in our case studies, this facilitates comparisons between traditional statistical tools like PCA and modern machine learning approaches such as autoencoders.
    \item The framework is accompanied by a user-friendly software implementation in our \proglang{R} package called \pkg{GLaRe} (Graphical Analysis of Latent Representations).
    The package provides graphical summaries to aid the selection of an optimal latent feature representation.
    It provides built-in latent feature representation methods as well as the option for the user to provide their own bespoke method.
\end{enumerate}
We demonstrate the practical utility of CLaRe through case studies on three high-dimensional datasets from diverse fields of application: 1) measurements of the mechanical strain at different locations in the eye, with application to the study of Glaucoma \parencite{lee_bayesian_2019}, 2) a neurobiological dataset of gel images of proteins from the brains of rats, with application to drug-use addiction \parencite{morris_pinnacle_2008}, and 3) the well-known MNIST dataset of handwritten digits that is widely used image recognition \parencite{lecun_mnist_1998}.
These case studies highlight that different datasets often favor different latent feature representation methods, emphasizing the importance of a consistent, objective and flexible evaluation framework.

The remainder of this article is structured as follows.
In Section \ref{sec:materials-and-methods} we present the methodological foundations of latent feature representations and generalization error that underpin CLaRe. 
In Section \ref{sec:case-studies}, we introduce three motivating datasets and employ the CLaRe framework to assess the performance of PCA, Discrete Wavelet Transform (DWT) and autoencoder representations of these datasets and select an optimal representation of each.
In Section \ref{sec:software}, we document the software implementation of our framework, GLaRe (Graphical Analysis of Latent Representations).
We close with a discussion in Section \ref{sec:discussion}.
\section{Methodological Framework (CLaRe)}\label{sec:materials-and-methods}

\subsection{High-Dimensional Object Data}

We use the term \emph{data object} to refer to the type and structure of the basic ``atom" of a statistical analysis
\parencite[][p.1]{marron_object_2021}.
In univariate statistics, the object is a number (i.e., scalar), and in classical multivariate statistics, observations comprise $p$ variables and are represented as $p$-dimensional vectors.
In areas such as genomics, advancements in data collection, storage and processing technologies mean that the data objects being collected are ultra high-dimensional vectors \parencite{stein_case_2010}.
There has also been an emergence of more general and complex data objects that vary over continua or grids, such as smooth time-varying curves \parencite{ramsay_functional_2005}, spiky signals \parencite{morris_wavelet-based_2006} and images \parencite{morris_automated_2011}, which when recorded at regular intervals also present as very high-dimensional vectors.
In most cases, a suitable transformation of the observed data objects to a lower-dimensional space of latent features, which we term a \emph{latent feature representation}, facilitates the application of familiar statistical approaches to the high-dimensional complex objects.

\subsection{Latent Feature Representations}

Suppose that we have $N$ observations of a data object, denoted by $X_1 (t), \dots, X_N(t)$, where $t$ indexes a location on a domain $\mathcal{T}$ over which the objects are defined.
For time-varying curves, $\mathcal{T}$ is generally a closed subset of the real line that represents a (normalized) time interval.
However, as exemplified in our three motivating datasets, the domain $\mathcal{T}$ can be multi-dimensional to represent locations in an image or surface, and it can also be non-Euclidean (e.g., our example Glaucoma data is defined on a partial spherical domain).
We assume that each observation is measured on a common\footnote{In practice, the measurement grids of individual observations need not be identical if they are all sufficiently fine such that interpolation onto a common, fine grid is feasible.}, ordered grid of $T$ points in $\mathcal{T}$, denoted by $\mathbf{t} = \left(t_1, \dots, t_T\right)^\top$, and we let $X_i(\mathbf{t}) = \left\{X_i(t_1), \dots, X_i(t_T)\right\}^\top$.
Then, we can represent the observed data in the $N \times T$ data matrix $\mathbf{X}$, which contains the vectors $X_1(\mathbf{t}), \dots, X_N(\mathbf{t})$ in its rows.
We refer to the $T$-dimensional space of features in which the observed data are represented as the \emph{data space}.

We define a \emph{latent feature representation method} as a technique comprising two transformations, known as the \emph{encoding} and \emph{decoding} transforms.
The encoding transform $f_{K}$ transforms an observation from the data space to a new space of latent features, called the \emph{representation space}
$$
f_{K} \left\{X_i(\mathbf{t})\right\} = \left(X_{i1}^*, \dots,  X_{iK}^* \right)^\top,
$$
where the number of features $K$ defines the dimensionality of the representation space and can range between $1$ and some possible maximum $K_{max}$. When $K \ll T$, we say that the latent feature representation is \emph{compact}.
As we expand on in Section \ref{sec:characterising-information-loss}, we typically want the representation space to be as compact as possible.

We define the decoding transform as the transformation function $g_K$ that maps an observation from the representation space back to the data space as
$$
\widehat{X}_i^{(K)} (\mathbf{t}) = g_K \left\{ \left(X_{i1}^*, \dots,  X_{iK}^* \right)^\top \right\}.
$$
In certain cases, $g_K = f^{-}_K$ when a generalized inverse $f^{-}_K$ exists.

Linear transformations of the form $f_{K} \left\{X_i(\mathbf{t})\right\} = \mathbf{A} X_i(\mathbf{t})$, for some $K \times T$ transformation matrix $\mathbf{A}$, are often used in practice.
For example, it is common to represent a functional observation $X_i(t)$ as a linear combination of a set of basis functions $\{\phi_k(t)\}_{k=1}^K$, which defines the decoding transformation
$$
\widehat{X}_i^{(K)} (\mathbf{t}) = \sum_{k=1}^K X_{ik}^* \phi_k(\mathbf{t}) = \boldsymbol{\Phi} \left(X_{i1}^*, \dots,  X_{iK}^* \right)^\top,
$$
where $\boldsymbol{\Phi} = \left[\phi_1(\mathbf{t}) | \dots | \phi_K(\mathbf{t}) \right]$ and the latent features $X_{ik}^*$ are basis coefficients. 
When these basis coefficients are computed by ordinary least squares, the encoding transformation $f_K$ is of the form $\mathbf{A} = \left( \boldsymbol{\Phi}^\top \boldsymbol{\Phi} \right)^{-1} \boldsymbol{\Phi}^\top$, and if $X_i(\mathbf{t})$ lies in the column space of $\boldsymbol{\Phi}$, then we have the inverse property $g_K=f^{-}_K$.
When the matrix of basis function evaluations $\boldsymbol{\Phi}$ is orthogonal, $\mathbf{A} = \boldsymbol{\Phi}^\top$, i.e., the transformation $f_K$ is simply right multiplication by this matrix.
However, in general, there is no need for the transformation $f_K$ to be orthogonal or even linear, and non-linear transformations may be preferred for certain types of data.

Although statistical modeling is performed in the representation space due to its attractive properties, we often want to transform modeling results back to the data space for inference, interpretation and visualization. 
As such, the accuracy and interpretation of an analysis depends on the degree of information that is preserved when moving back and forth between the data and representation spaces for a given latent feature representation.
In what follows, we characterize the degree of information loss of a latent feature representation on a full dataset.

\subsection{Characterizing the Full Distribution of Information Loss}\label{sec:characterising-information-loss}

We denote the degree of information loss of a latent feature representation for each individual observation by 
$$
\text{Loss} \left\{ X_i(\mathbf{t}), \widehat{X}_i^{(K)}(\mathbf{t}) \right\},
$$
where the $\text{Loss}(\cdot, \cdot)$ is a symmetric, non-negative function that measures dissimilarity between $X_i(\mathbf{t})$ and $\widehat{X}_i^{(K)}(\mathbf{t})$, and satisfies $\text{Loss}\left\{ X_i(\mathbf{t}), \widehat{X}_i^{(K)}(\mathbf{t}) \right\}=0$ when $X_i(\mathbf{t}) \equiv \widehat{X}_i^{(K)}(\mathbf{t})$.
For example, $\text{Loss}(\cdot, \cdot)$ could be the Euclidean distance between $X_i(\mathbf{t})$ and $\widehat{X}_i^{(K)}(\mathbf{t})$ \parencite{morris_comparison_2017}, or the complement of a similarity measure such as a squared correlation or concordance index \parencite{yang_quantile_2020}.
We say that our latent feature representation \emph{lossless} for the $i$th observation if
$$
\text{Loss} \left\{X_i(\mathbf{t}), \widehat{X}_i^{(K)}(\mathbf{t}) \right\} = 0,
$$
and lossless for the full dataset if
$$
\text{Loss} \left\{ X_i(\mathbf{t}), \widehat{X}_i^{(K)}(\mathbf{t}) \right\} = 0 \quad \forall \quad  i = 1, \dots, N.
$$
That is, we only refer to a latent feature representation as lossless for a given dataset if the representation is lossless for every individual observation in that dataset.
More generally, we can allow some tolerance of information loss and say that
the transformation $f_K$ is \emph{near-lossless} for the $i$th observation if
$$
\text{Loss} \left\{ X_i(\mathbf{t}), \widehat{X}_i^{(K)}(\mathbf{t}) \right\} < \epsilon,
$$
for a chosen tolerance level $\epsilon$. 
Similarly, we say that the transformation $f_K$ is near-lossless for the full dataset only if each individual observation achieves this tolerance level, that is
$$
\text{Loss} \left\{ X_i(\mathbf{t}), \widehat{X}_i^{(K)}(\mathbf{t}) \right\} < \epsilon \quad \forall \quad  i = 1, \dots, N.
$$
It is important to distinguish this definition from an overall (or total) measure such as the average of individual losses $\frac{1}{N}\sum_{i=1}^N \text{Loss} \left\{ X_i(\mathbf{t}), \widehat{X}_i^{(K)}(\mathbf{t}) \right\}$.

\begin{figure}
    \centering
    \includegraphics[width=0.75\linewidth]{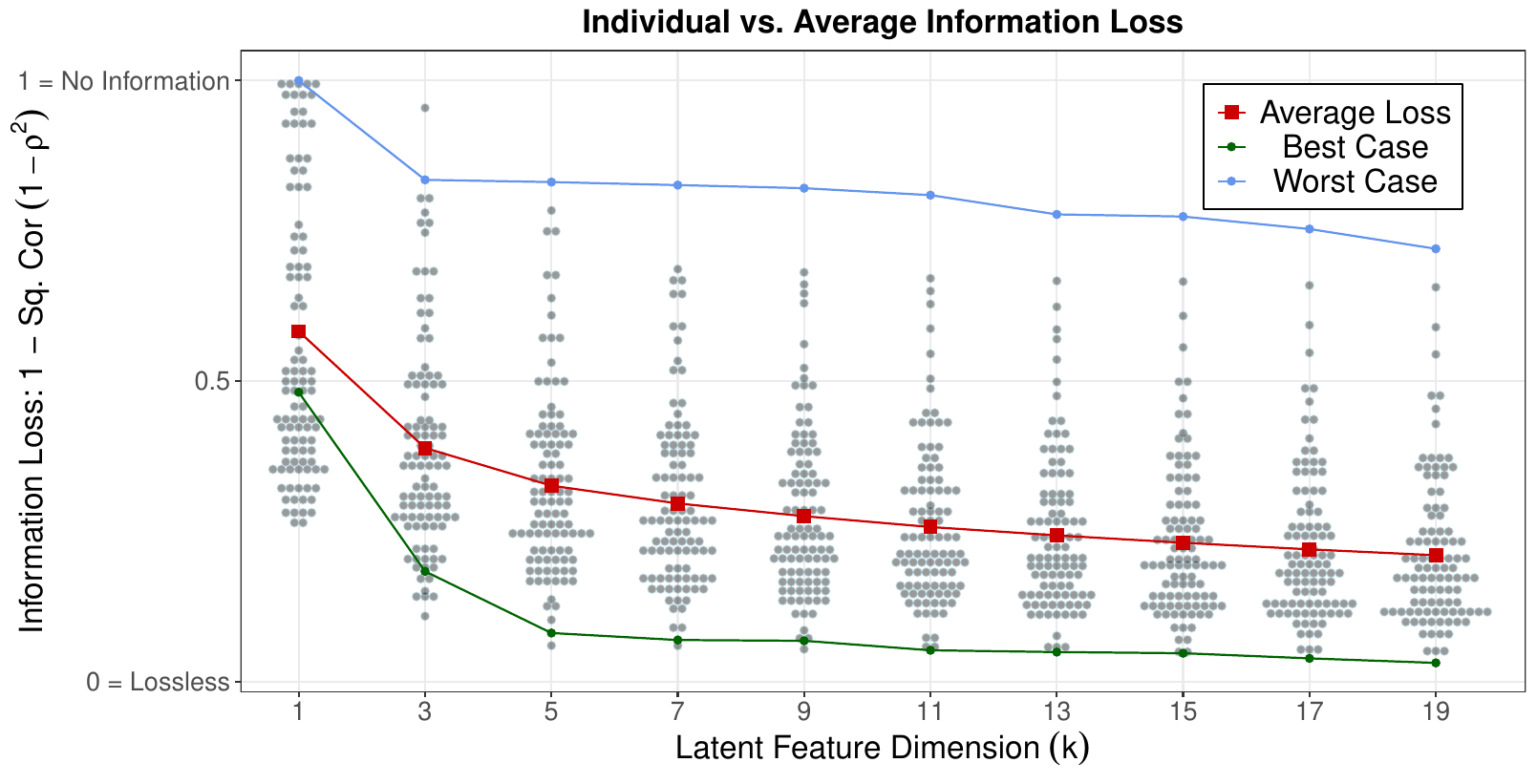}
    \caption{
    Generalization errors from a PCA representation of the \texttt{phoneme} data (see Appendix \ref{sec:additional-data}). The grey dots represent individual cross-validated reconstruction errors from PCA representations using different numbers of latent features ranging from $1$ to $19$ ($x$ axis).
    The complement of the squared correlation is used as reconstruction loss, so $1$ indicates no information retained by the representation and $0$ indicates a lossless representation ($y$ axis).
    The square red points and solid red line trace the average reconstruction error. The green and blue points / lines trace the performance of the best and worst performing observations (selected at $K=19$), respectively.
    }
    \label{fig:ind-losses}
\end{figure}

Figure \ref{fig:ind-losses} displays the distribution of individual information losses from applying PCA with varying latent dimensions to the \texttt{phoneme} dataset (a dataset of $1$-dimensional signals, described in more detail in Appendix \ref{sec:additional-data}).
In this case, we are using the complement of the squared correlation as our loss, so a value of $1$ means the representation captures no information and a value of $0$ means that the representation is lossless.
The gray points represent the individual observations' losses, whereas the red squares indicate the average loss.
The figure highlights the information that can be hidden when a single statistic is used to describe the full distribution of losses. For example, at $k = 1$, the average loss is at approximately $0.6$ but there are observations with individual losses at almost $1$, which would indicate that no information is being retained by the transformation.

Achieving a chosen tolerance level for every observation (i.e., requiring that even the worst case meets the near-losslessness threshold) can sometimes be unrealistic in practice.
Often, a small number of observations (e.g., the one traced by the blue points and line in Figure \ref{fig:ind-losses}) possess idiosyncratic features that cannot be captured by a representation that is otherwise compact and near-lossless for the vast majority of the observations.
In this case, using the worst case may be overly stringent and result in a latent feature representation that is higher-dimensional and more complex than required for the vast majority of observations.
Therefore, we generalize the notion of near losslessness for the entire dataset, and require it to be met for \emph{quantiles} or \emph{percentiles} of the distribution of individual generalization errors, in which the worst-case observation is the $100$th percentile. 
As before, $\epsilon$ denotes the tolerance level of information loss that we want to achieve.
We now introduce the \emph{attainment rate}, which we denote by $1 - \alpha$, as the proportion of observations that we want to achieve this tolerance level.
For example, an attainment rate of $1 - \alpha = 0.95$ would indicate that we require $95\%$ of the observations in the dataset to achieve a loss smaller than the tolerance level $\epsilon$.
We refer to a chosen combination of $\epsilon$ and $(1 - \alpha)$ as the qualifying criterion and we use the term \emph{qualifying dimension (qd)} to denote the smallest latent feature dimension $K$ for which this criterion is satisfied.
When comparing two latent feature representation methods (e.g., PCA and autoencoder), for a fixed $\epsilon$ and $\alpha$, we generally prefer the method with a smaller qualifying dimension as it provides a more compact representation of the dataset, although other properties might also be taken into account depending on the setting, as we discuss later.
Finally, we characterize the dimension reduction achieved by a method by its \emph{compression ratio}, which is the ratio of the original data dimension $T$ to the qualifying dimension $qd$, typically rounded to the nearest whole number.

\begin{figure}
    \centering
    \includegraphics[width=1\linewidth]{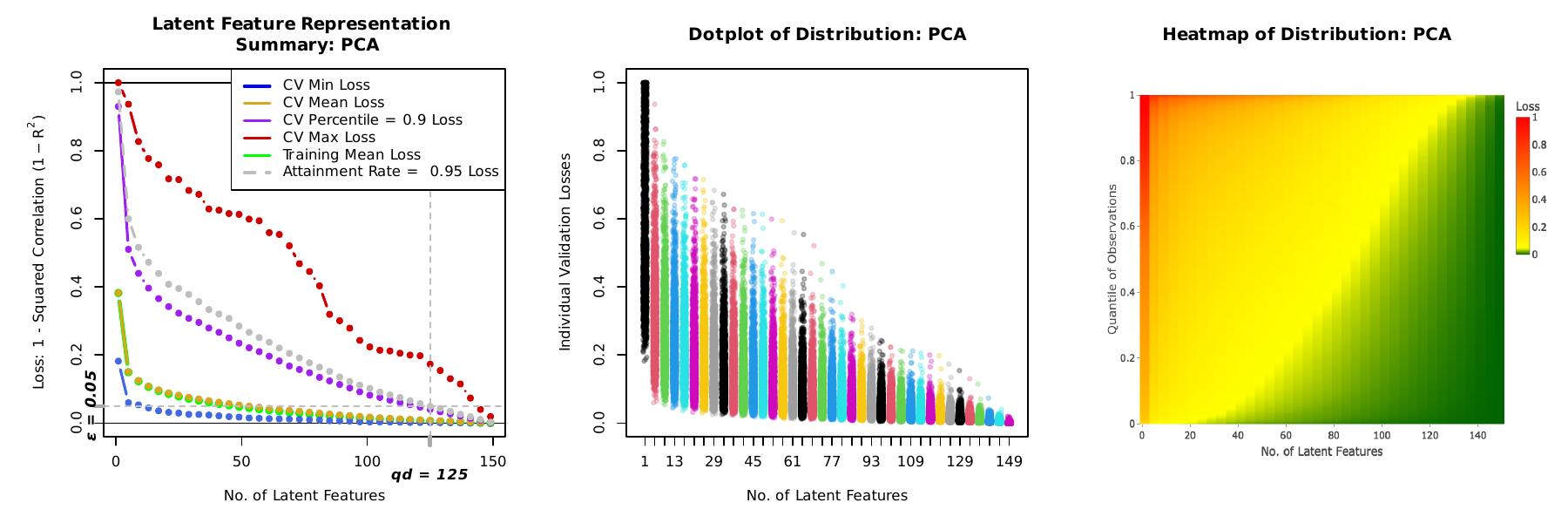}
    \caption{Three options for displaying the full distribution of information losses, demonstrated for a PCA representation of the \texttt{phoneme} data (see Appendix \ref{sec:additional-data}).
    \textbf{Left}: The summary plot returned by default by our GLaRe software, which summarizes the full distribution of cross-validated information losses by displaying a number of its quantiles.
    \textbf{Middle}: A jittered dot-plot of the full distribution of information losses.
    \textbf{Right}: A heatmap of the full distribution of information losses.
    }
    \label{fig:viz-options}
\end{figure}

There are a number of ways to visualize the full distribution of information losses (Figure \ref{fig:viz-options}). 
The left-hand panel of Figure \ref{fig:viz-options} displays the summary graphic that we have designed to accompany our CLaRe framework and is returned by default the \texttt{GLaRe()} function in our software implementation, GLaRe.
The overall cross-validated loss is displayed in yellow, with the analogous loss computed on the training data shown in green for comparison.
Different quantiles of the distribution of individual cross-validated losses are displayed to summarize the full distribution: the minimum and maximum are shown in blue and red, respectively, a user-specified quantile of the distribution ($0.9$ in this example) is displayed in purple and the quantile of the distribution being used as the attainment rate $1 - \alpha$ (defaulting to $0.95$, i.e., the $95$th percentile) is displayed in light gray.
The corresponding value of the tolerance level $\epsilon$ is overlaid as a gray dashed horizontal line, and hence the latent feature dimension (i.e., location on the $x$-axis) at which the two gray lines meet corresponds to the qualifying dimension.
The tolerance level $\epsilon$ and the qualifying dimension ($qd$) are marked in bold and italic typeface on the $y$ and $x$ axes, respectively.

The middle and right-hand panels of Figure \ref{fig:viz-options} display alternative options for visualizing the same distribution.
The middle panel displays a jittered dot-plot, where each point represents an individual value of the information loss distribution and the points are colored according to the latent feature dimension $K$.
The right-hand panel presents a heatmap to display of the full distribution of information losses.
The latent feature dimension is represented on the $x$-axis, the corresponding quantile of the information loss distribution at that feature dimension is shown on the $y$-axis and the color indicates the value of the information loss at that feature dimension and quantile.
Hereafter, we will use the plot in left-hand panel of Figure \ref{fig:viz-options} to present the results of applying our framework to different datasets, but all three visualization options are available in our software implementation.

\subsection{Cross-validated Estimation of Information Loss}

Estimating a latent feature representation can equivalently be viewed as a prediction problem, where the goal is to construct $f_{K}$ and $g_K$ such that the predictions $\widehat{X}_i^{(K)}(\mathbf{t})$ match the observed data $X_i(\mathbf{t})$ as closely as possible \parencite{krzanowski_cross-validation_1987, wold_cross-validatory_1978, diana_cross-validation_2002, bro_cross-validation_2008}.
Hence, it is possible that a latent feature representation method can \emph{overfit} to the data on which it is being learned.
If the information loss of a latent feature representation method is evaluated on that same dataset, then it will tend to be overly optimistic and will not accurately quantify the method's true \emph{generalization error}, which we define as its information loss in reconstructing unseen data.
To obtain a valid estimate of generalization error, it is necessary to use an independent \emph{validation} dataset that is not used to learn the latent feature representation \parencite{diana_cross-validation_2002, bro_cross-validation_2008}.

Generally, with limited data, it is inefficient to perform a single split of the dataset into training and validation sets, as a single random split will tend to be variable, i.e., if the split were performed again, the new validation set would produce a different estimate of information loss \parencite[Table 1]{collins_evaluation_2024}.
Additionally, because we are interested in the individual values of information loss, rather than an average, sample splitting would only provide us with these values for observations included in the validation set. 
To mitigate these concerns, we employ \emph{cross-validation} \parencite{stone_cross-validatory_1974}, where the data are systematically divided into different training and validation splits, called folds, and the training and validation is performed separately for each split.
When we are interested in an average or total estimate of information loss, cross-validation will tend to be more stable than sample splitting because the estimates are averaged over different folds.
For our purposes, it additionally produces a generalization error estimate for each individual observation in the dataset.
Although cross-validation has long been understood as necessary to estimate generalization error for latent feature representation methods, in particular PCA (e.g., dating back to the work of \textcite{wold_cross-validatory_1978, eastment_cross-validatory_1982, krzanowski_cross-validation_1987}), it is not automatically returned by standard software packages or routinely used in practice to choose between different methods.
Algorithm 1 provides a high-level overview of the full CLaRe framework for evaluating latent feature representation methods.


\begin{algorithm}
\caption{CLaRe Framework for Evaluating Latent Representations}
\begin{algorithmic}[1]

\State \textbf{Initialization} Start with a data matrix $\mathbf{X}$, a latent feature representation method (e.g., PCA, wavelets, or autoencoders) and a choice of loss function. Define the range of latent dimensions to evaluate and the number of folds for cross-validation. Set up a matrix to hold the cross-validated information losses for all observations (in rows) and latent dimensions (in columns).

\State \textbf{Generate cross-validation splits:} Randomly shuffle the dataset and then divide it into folds for cross-validation. Reuse these splits across all candidate latent feature dimensions for consistency and efficiency.

\State \textbf{For each candidate latent dimension $K$:}
\begin{enumerate}[a.]
    \item \textbf{For each cross-validation fold:}
    \begin{enumerate}[i.]
        \item \textbf{Split into Training/Validation datasets:}
        \begin{itemize}
            \item Split the data into training and validation sets for the current fold.
        \end{itemize}
        \item \textbf{Learn encoding and decoding transformations on training data:}
        \begin{itemize}
            \item Train a representation method (e.g., PCA, wavelets, or an autoencoder) on the training data to learn the encoding and decoding transformations $f_K$ and $g_K$.
        \end{itemize}
        \item \textbf{Reconstruct validation data:}
        \begin{itemize}
            \item Apply the learned transformations to encode and decode the validation dataset to give reconstructions $\widehat{X}_i (\mathbf{t})$ of $X_i (\mathbf{t})$.
        \end{itemize}
        \item \textbf{Compute information loss on validation data:}
        \begin{itemize}
            \item Measure the dissimilarity between the original and reconstructed validation data for each observation in the validation set using the loss function $\text{Loss}(\cdot, \cdot)$ and store these values.
        \end{itemize}
    \end{enumerate}
\end{enumerate}

\State \textbf{Identify the qualifying dimension:} For each latent dimension, compute the $1 - \alpha$th quantile (e.g., 95th percentile) of the information loss across all observations as specified by the attainment rate $1 - \alpha$. Select the smallest latent feature dimension where this quantile is below the specified tolerance level $\epsilon$.

\State \textbf{Fit the model at the qualifying dimension:} If a qualifying dimension is identified, fit the representation method using the full dataset at the qualifying dimension. Store the final trained model for downstream applications.

\State \textbf{Return results:} Return the full matrix of cross-validated information losses and, if applicable, the qualifying latent dimension and the final trained model at that dimension. If applying the algorithm to select among different methods, choose the method with the smallest qualifying dimension.

\end{algorithmic}
\end{algorithm}

\section{Case Studies}\label{sec:case-studies}

\begin{figure}
    \centering
    \includegraphics[width=1\linewidth]{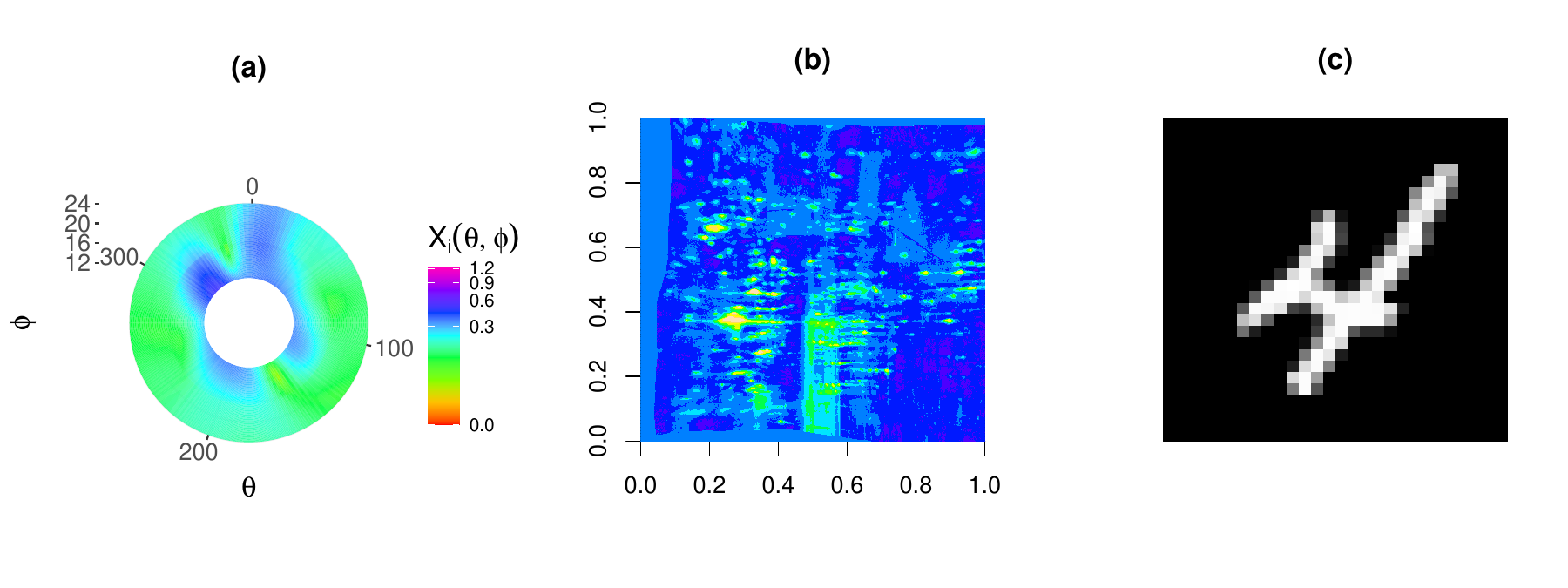}
    \caption{
    A sample observation from each of our three motivating datasets.
    \textbf{(a)}: A sample glaucoma image, representing a polar azimuthal projection of MPS functions for a single eye at one IOP level. A cubed-root transformation is applied for visualization.
    \textbf{(b)}: A sample 2D gel electrophoresis image, showing proteomic content in the brain tissue of a rat.
    \textbf{(c)}: A sample MNIST digit image, which is a $28 \times 28$ pixel greyscale image of a single handwritten digit.}
    \label{fig:combined-data-objects}
\end{figure}

In this section, we introduce three motivating datasets -- the Glaucoma dataset, the Proteomic Gels dataset and the MNIST digits dataset (Figure \ref{fig:combined-data-objects} (a)--(c)).
For each dataset, we apply our CLaRe framework to select among the following three latent feature representation methods\footnote{Specific implementation details for the methods are described in Section \ref{sec:software}.}:
\begin{enumerate}
    \item \textbf{Principal Components Analysis (PCA)} \parencite{hotelling_analysis_1933} is the most prevalent dimension reduction technique in statistics, used to represent multivariate data in a lower-dimensional space that preserves as much variance as possible. 
    Using terminology from Section \ref{sec:materials-and-methods}, the encoding transform in PCA is the linear projection of the data matrix $\mathbf{X}$ onto the matrix containing the leading $K$ eigenvectors of the empirical covariance matrix, typically denoted by $\mathbf{U}$, which yields the latent features, known as PCA scores.
    The decoding transformation is the inverse linear mapping of the latent features back to the data space, which consists of multiplying the latent features by the transposed matrix of eigenvectors.
    \item \textbf{Thresholded Discrete Wavelet Transform (DWT)}: The DWT \parencite{daubechies_wavelet_1990, daubechies_ten_1992} is a dimension reduction technique that linearly transforms a discretely-sampled signal of length $T$ into a set of $T$ latent features, known as wavelet coefficients, that are localized in both time and frequency. In theory, both the DWT and the inverse DWT (IDWT) are linear transformations but they benefit from efficient recursive algorithms in practice  \parencite{daubechies_ten_1992}.
    
    To use the DWT as a dimension reduction technique, we exploit its inherent sparsity property, i.e., that a signal can typically be well represented by $K \ll T$ non-zero wavelet coefficients. We apply the DWT to each observation and then learn the $K$ most important wavelet coefficients jointly across all observations (see Appendix \ref{sec:wavelet-thresholding-algorithm} for more details) -- these coefficients are retained as the $K$ latent features. 
    The remaining latent features are set to $0$ when applying the IDWT as the decoding transform. 
    Thus, encoding consists of the composition of two transformations: (1) the linear DWT transformation and (2) the nonlinear selection of the $K$ most important coefficients while zeroing the rest.
    The thresholded DWT has been used in functional data analysis \parencite[see, e.g.,][]{morris_automated_2011} and for simplicity we abbreviate as DWT hereafter.

    \item \textbf{An Autoencoder (AE)} \parencite{rumelhart_learning_1986} is a dimension reduction technique from the field of machine learning.
    The encoding and decoding transformations that map the data to the latent representation space, known as the \emph{bottleneck}, are general non-linear functions parameterized by neural networks
    $$
    f_K(X_i) = \text{NN}_{\boldsymbol{\theta}_E}(X_i (\mathbf{t})) \text{ and } g_K\big\{\left(X_{i1}^*, \dots,  X_{iK}^* \right)^\top\big\} = \text{NN}_{\boldsymbol{\theta}_D} \big\{\left(X_{i1}^*, \dots,  X_{iK}^* \right)^\top\big\},
    $$
    where $\boldsymbol{\theta}_E$ and $\boldsymbol{\theta}_D$ are vectors of weights for the encoding and decoding networks with dimensions determined by the respective neural network architectures.
    These weights are learned to minimize the reconstruction error
    $$
    \sum_{i=1}^N \bigg\| X_i (\mathbf{t}) - \widehat{X}_i^{(K)}(\mathbf{t}) \bigg\|^2,
    $$
    or another suitable loss measure for the problem at hand \parencite[e.g., binary cross-entropy for binary data, see][]{kalinowski_keras_2024}.
\end{enumerate}
The three methods originate from different fields -- PCA is a conventional statistical method, the DWT originates from signal processing and autoencoders are ubiquitous in machine learning and artificial intelligence.
They also exhibit differences in the amount of structure they learn from the data.
On one hand, the thresholded DWT learns the least structure from the data -- the linear DWT and IDWT transforms are fixed and only the ordering of the coefficients to threshold is learned from the data.
PCA is more flexible and learns more structure from the data because the matrix of eigenvectors $\mathbf{U}$ that defines the encoding and decoding transformations is computed from the empirical covariance matrix of the data.
An AE is even more flexible and learns greater structure from the data, because the linearity restriction in PCA is relaxed and the AE learns general non-linear encoding and decoding transformations.

To explore the practical implications of these differences and how they relate to sample size, conclude in Section \ref{sec:sample-size-experiment} by presenting the results of an experiment where we artificially decimate the sample size of the Glaucoma dataset.
We demonstrate that more flexible empirical methods (e.g., PCA) work well when there is enough data to reliably estimate the latent features, and that fixed methods (e.g., DWT) perform better when there is not enough data.
For all applications in this section, we use a tolerance level of $\epsilon = 0.05$, an attainment rate of $1 - \alpha=0.95$ and employ the complement of the squared correlation as our loss function.

\subsection{Glaucoma Data}

Glaucoma is considered a leading factor in blindness. It is characterized by damage to the optic nerve, which can be induced by \emph{intraocular pressure (IOP)}. 
To investigate proposed hypotheses about the relationship between glaucoma and IOP, \textcite{fazio_age-related_2014} developed instrumentation to measure the mechanical strain on the scleral surface of the eye at different levels of IOP. 
The measurements were summarized at different locations on the scleral surface of the eye as \emph{maximum principal strain} (MPS). MPS was computed on 34 eyes from 19 normal human donors. It was measured in the posterior globe of both eyes on a partial spherical domain with $120$ circumferential locations $\upsilon \in (0^{\circ}, 360^{\circ})$ and $120$ meridional locations $\theta \in (9^{\circ}, 24^{\circ})$.
These measurements were taken at 9 different IOP levels.
One study goal for this dataset was to test the hypothesis that scleral strain decreases with age thereby leaving the optic nerve head susceptible to damage which could be a contributing factor in the development of glaucoma \parencite{lee_bayesian_2019}.
Figure \ref{fig:combined-data-objects} (a) displays a two-dimensional polar azimuthal projection of a single observation from the Glaucoma data.

We let $X_i(t)$ denote the MPS function for a single eye at a specific IOP level so that $i = 1, \dots, N = 306$ ($34$ eyes at $9$ IOP levels).
The data lives on a domain $\mathcal{T}$ which is the portion of the sphere defined by $(\upsilon, \theta)$ for $\upsilon \in (0^{\circ}, 360^{\circ})$ and $\theta \in (9^{\circ}, 24^{\circ})$.
Therefore, each observation $X_i(t)$ is recorded on a common grid of size $T = 14400$.
The recordings are indexed by the $14400$-dimensional vector $\mathbf{t} = \boldsymbol{\upsilon} \times \boldsymbol{\theta}$, where $\boldsymbol{\upsilon}$ represents $120$ equally-spaced measurements of $\upsilon$ along $(0^{\circ}, 360^{\circ})$ and $\boldsymbol{\theta}$ represents $120$ equally-spaced measurements of $\theta$ along $(9^{\circ}, 24^{\circ})$.
We denote the vector of measurements for the $i$th observation as $X_i(\mathbf{t})$, so that the full dataset can be represented by the $N \times T$ data matrix $\mathbf{X}$, containing $X_1(\mathbf{t}), \dots, X_N(\mathbf{t})$ in its rows.

\begin{figure}
    \centering
    \includegraphics[width=1\linewidth]{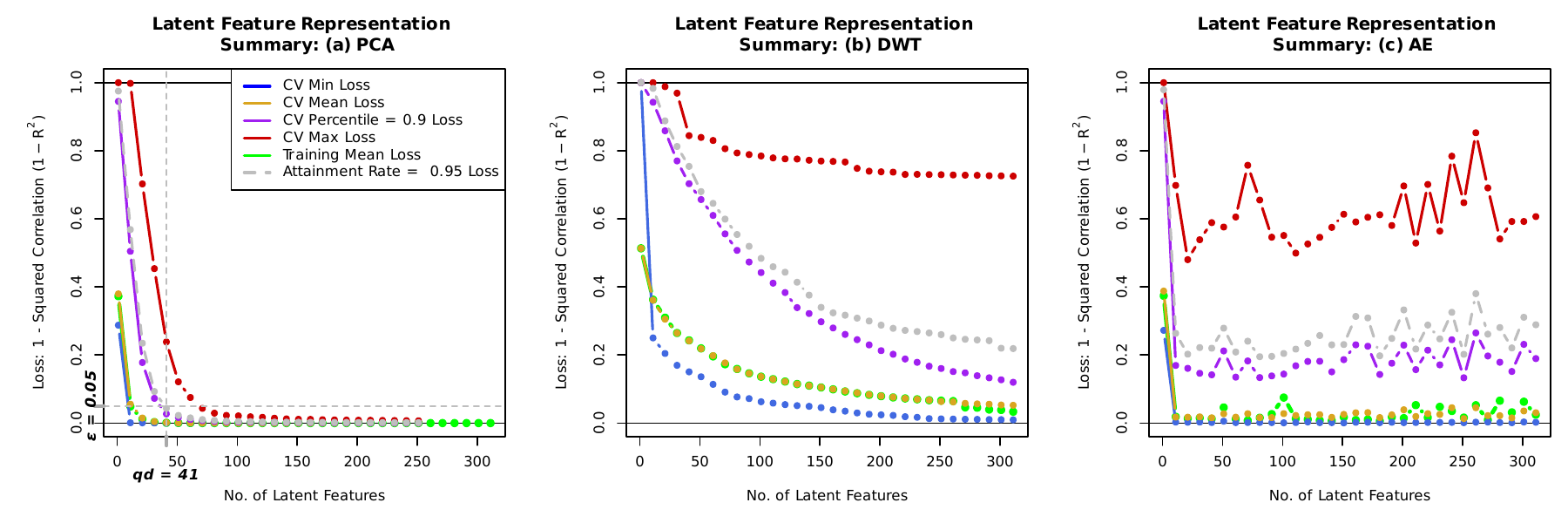}
    \caption{Summary \texttt{GLaRe()} plot for the Glaucoma data. A grid of equally-spaced values from $1$ to $301$ in increments of $10$ was used for the latent feature dimensions. 
    The legend for plots (b) and (c) has been suppressed so that it does not mask the worst-case observation and because it is identical to the legend for plot (a).}
    \label{fig:eye-results}
\end{figure}

Figure \ref{fig:eye-results} displays the summary plot from the application of \texttt{GLaRe()} to the Glaucoma data.
PCA is the most suitable latent feature representation method for this dataset because it achieves the qualifying criterion with qualifying dimension $qd=41$, whereas DWT and AE do not achieve the qualifying criterion for $K \leq 301$.
A grid of equally-spaced values from $1$ to $301$ in increments of $10$ was used for the latent feature dimensions.
Although it was possible to use larger latent feature dimensions for the DWT and AE, it was deemed unnecessary because the qualifying criterion was achieved for PCA with a qualifying dimension $qd=41$ and is the favored method for this dataset.
PCA achieved a compression ratio of $351:1$ ($T = 14400$ to $qd = 41$).
The computation times for PCA, DWT and AE were $1.6$, $1.7$ and $67.4$ minutes, respectively.

\subsection{Proteomic Gels Data}

In neurobiology, a particularly important issue is the identification of changes responsible for the transition from non-dependent drug use to addiction which is characterized by drug intake behavior. Studies on rats have shown that rats given a 6-12 hours/day access to cocaine or heroin have significant increase in drug intake while rats given 1 hour/day access kept the same level of intake over time.
The corresponding neurochemical changes in the extended part of the brain amygdala relate to cellular effects that affect protein expression and function which can be detected via proteomic analysis. To study this phenomenon, experiments were conducted in which the rats were trained to get cocaine by pressing a lever: 6 rats were given 1hour/day access, 7 rats were given 6 hours /day access and 8 rats were used for control with no access to cocaine. The rats were euthanized after some time, and their brains studied \parencite{morris_pinnacle_2008}. 
Two-dimensional gel electrophoresis was used to study the proteomic content in the brain tissues. 
Between two and three gels were obtained from each rat and brain region, resulting in a dataset of 53 gel images from 21 rats.
Each gel image has $556,206$ pixel intensities observed on a $646 \times 861$ grid. 
A research goal for this dataset was to study the proteins which are differentially expressed in the brains of rats that were exposed to cocaine for a long time versus those that were not.
This can be done by finding regions in the gel images where image intensity is significantly different across groups \parencite{morris_statistical_2012}. 
Figure \ref{fig:combined-data-objects} (b) displays a single gel image observation.

We denote a single observation from the Proteomic Gels dataset as $X_i(t)$, where $i=1, \dots, N = 53$.
Here, $t$ represents a location $(t_1, t_2)$ in the two-dimensional Euclidean domain defined by the Cartesian product $\mathcal{T} = [0, 1] \times [0, 1]$.
Measurements of each observation are made at the vector of locations $\mathbf{t} = \mathbf{t}_1 \times \mathbf{t}_2$ where $\mathbf{t}_1$ and $\mathbf{t}_2$ represent vectors of $646$ and $861$ equally-spaced points along $[0, 1]$, respectively.
Then $X_i (\mathbf{t})$ denotes the $T = 556206 ( = 646 \times 861)$-dimensional vector containing the measurements of the $i$th observation at the locations in $\mathbf{t}$, and the $N \times T$ data matrix $\mathbf{X}$ contains $X_1 (\mathbf{t}), \dots, X_N (\mathbf{t})$ in its rows.

\begin{figure}
    \centering
    \includegraphics[width=1\linewidth]{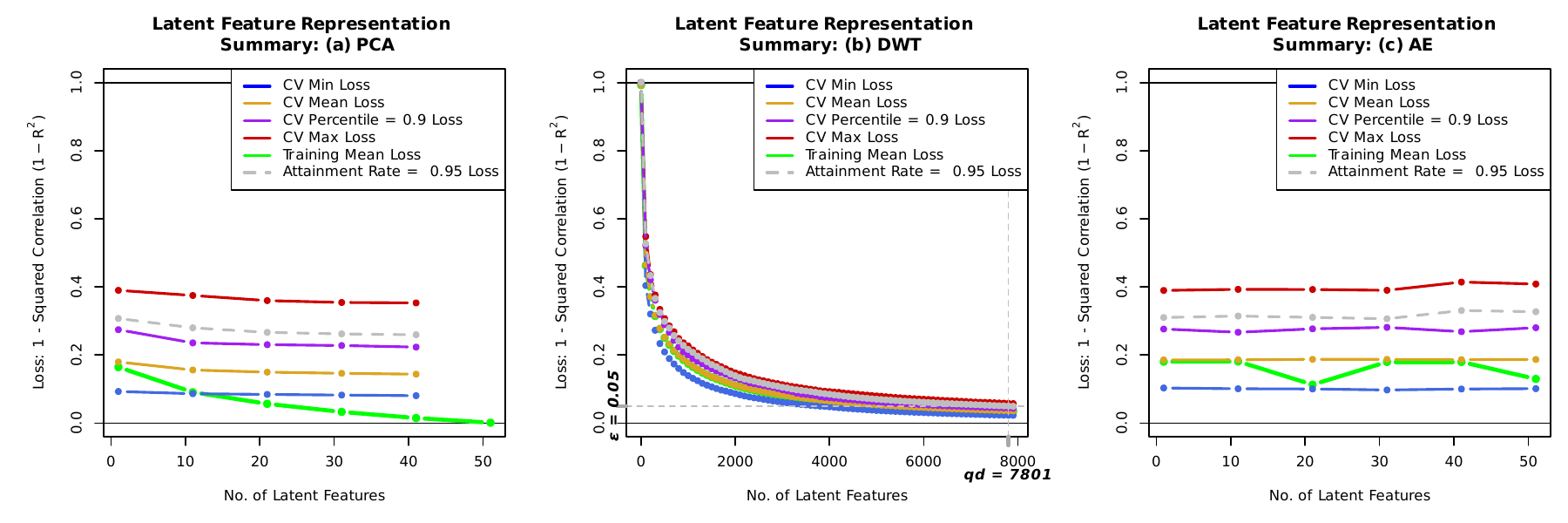}
    \caption{Summary \texttt{GLaRe()} plot for the Proteomic Gels data. For PCA and AE, a grid of equally-spaced values from $1$ to $53$ in increments of $10$ was used for the latent feature dimensions.
    For the DWT,  a grid of equally-spaced values from $1$ to $8000$ in increments of $10$ was used for the latent feature dimensions.}
    \label{fig:gels-results}
\end{figure}

Figure \ref{fig:gels-results} displays the summary plot from the application of \texttt{GLaRe()} to the Proteomic Gels data.
Because this dataset only contains $N=53$ observations, the maximum possible latent dimension for PCA is $52$ and PCA does not achieve the qualifying criterion for $K \leq 52$.
Due to computational constraints, we ran the AE for the same range of candidate latent feature dimensions, and it did not achieve the qualifying criterion.
Performance did not appear to improve as the latent feature dimension was increased.
In contrast, the maximum latent dimension for the DWT is not constrained and hence we ran \texttt{GLaRe()} on a grid of equally-spaced values from $1$ to $8000$ in increments of $100$.
The qualifying dimension for the DWT is $qd=7801$.
Hence, the DWT is the favored representation method for the Proteomic Gels dataset and it provides a compression ratio of $71:1$ ($T = 556206$ to $qd = 7801$).
The computation times for PCA, DWT and AE were $0.9$, $47.6$ and $109.6$ minutes, respectively.

\subsection{MNIST Digits Data}

The MNIST (Modified National Institute of Standards and Technology) database of handwritten digits was compiled by \textcite{lecun_mnist_1998}, from a larger collection of images from the National Institute of Standards and Technology (NIST).
It comprises a training set of $60000$ images and a test set of $10000$ images, representing hand-written digits from $0$ to $9$ (i.e., $10$ distinct digits/classes).
The original black and white images from NIST were modified into $28 \times 28$ pixel greyscale images.
The MNIST dataset has been employed extensively in computer vision and deep learning applications as a test case for image reconstruction and digit identification/ classification models.
The dataset can be represented in an $N \times T$ matrix $\mathbf{X}$, where $N = 60000$ (in the case of the training set) and $T= 784$ $(= 28 \times 28)$.
We let $X_i(t)$ represent the value of the $i$th greyscale image at pixel location $t$, where $t \in \mathbf{t} = \{1, \dots, 28\} \times \{1, \dots, 28\}$.
Then the $i$th row of the data matrix $\mathbf{X}$ contains the $784$-dimensional vector $X_i(\mathbf{t})$, i.e., measurements of the $i$th observation at the vector of pixel locations in $\mathbf{t}$.
We normalise the greyscale images so that $X_i(t) \in [0, 1]$ for all $t \in \mathbf{t}$ and $i = 1, \dots, N$. 
Figure \ref{fig:combined-data-objects} (c) displays a single digit from the MNIST dataset.
The full dataset (training and test) is publicly available in the \pkg{keras} \proglang{R} package \parencite{kalinowski_keras_2024} and can be loaded using the \texttt{dataset\_mnist()} function.

\begin{figure}
    \centering
    \includegraphics[width=1\linewidth]{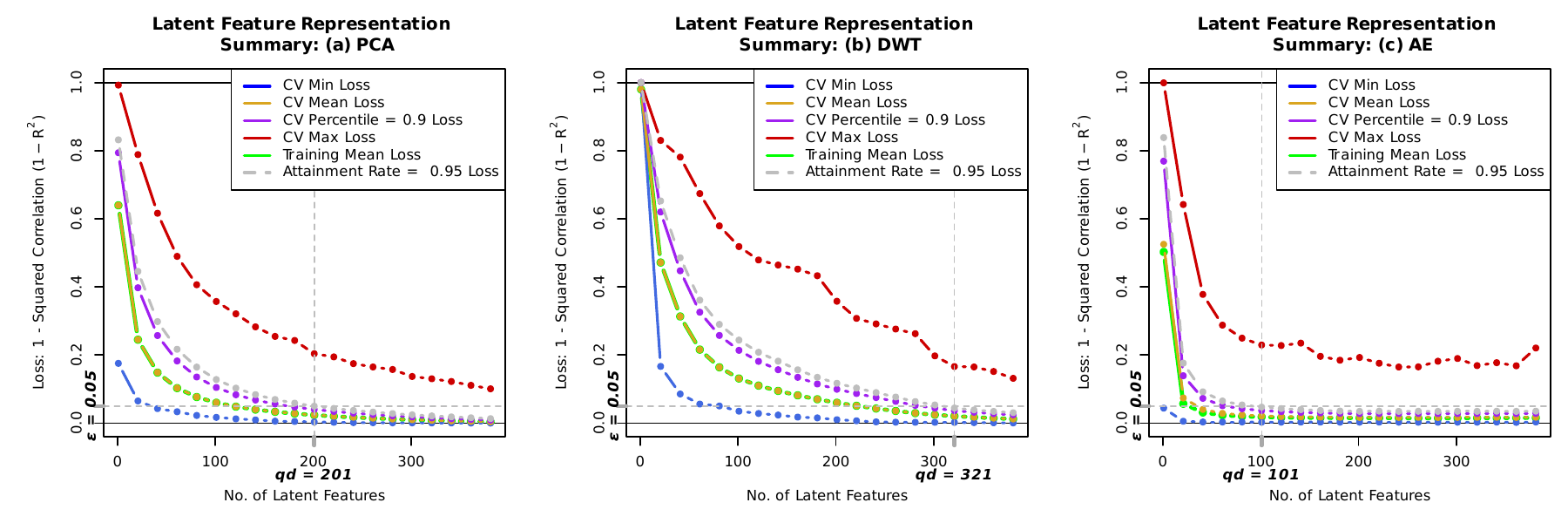}
    \caption{Summary \texttt{GLaRe()} plot for the MNIST data. A grid of equally-spaced values from $1$ to $381$ in increments of $20$ was used for the latent feature dimensions.}
    \label{fig:mnist-results}
\end{figure}

Figure \ref{fig:mnist-results} displays the summary plot from the application of \texttt{GLaRe()} to the MNIST data.
A grid of equally-spaced values from $1$ to $381$ in increments of $20$ was used for the latent feature dimensions.
All three latent feature representation methods achieve the qualifying criterion within this range: $qd = 201$ for PCA, $qd = 321$ for the DWT and $qd = 101$ for the AE.
Hence, the AE is the preferred method for this dataset because it provides the most compact (i.e., smallest $qd$) representation.
The AE has a compression ratio of $8:1$ ($T = 784$ to $K = 101$).
The computation times for PCA, DWT and AE were $16.2$, $27.4$ and $1115.3$ minutes, respectively.

\subsection{Sample Size Experiment}\label{sec:sample-size-experiment}

We  present the results of an experiment that demonstrates the dependence of flexible latent feature representation methods such as PCA and AE on sample size.
In PCA, the encoding and decoding transformations are learned entirely from the data, so it is highly dependent on having a sufficient sample size.
In contrast, the DWT transformation is fixed a-priori and only the ordering of the wavelet coefficients to retain is learned from the data and hence there is less reliance on sample size.
We demonstrate this concept empirically on the Glaucoma dataset. 
We start with the full dataset ($N=306$) and then sub-sample the dataset to create smaller datasets of sizes $N=153$, $N=76$ and $N=38$ respectively.
We run \texttt{GLaRe()} to compare the performance of PCA and the thresholded DWT as the sample size is successively degraded.
In all cases, because of the small sample sizes, we use leave-one-out rather than $k$-fold cross-validation.

\begin{figure}
    \centering
    \includegraphics[width=1\linewidth]{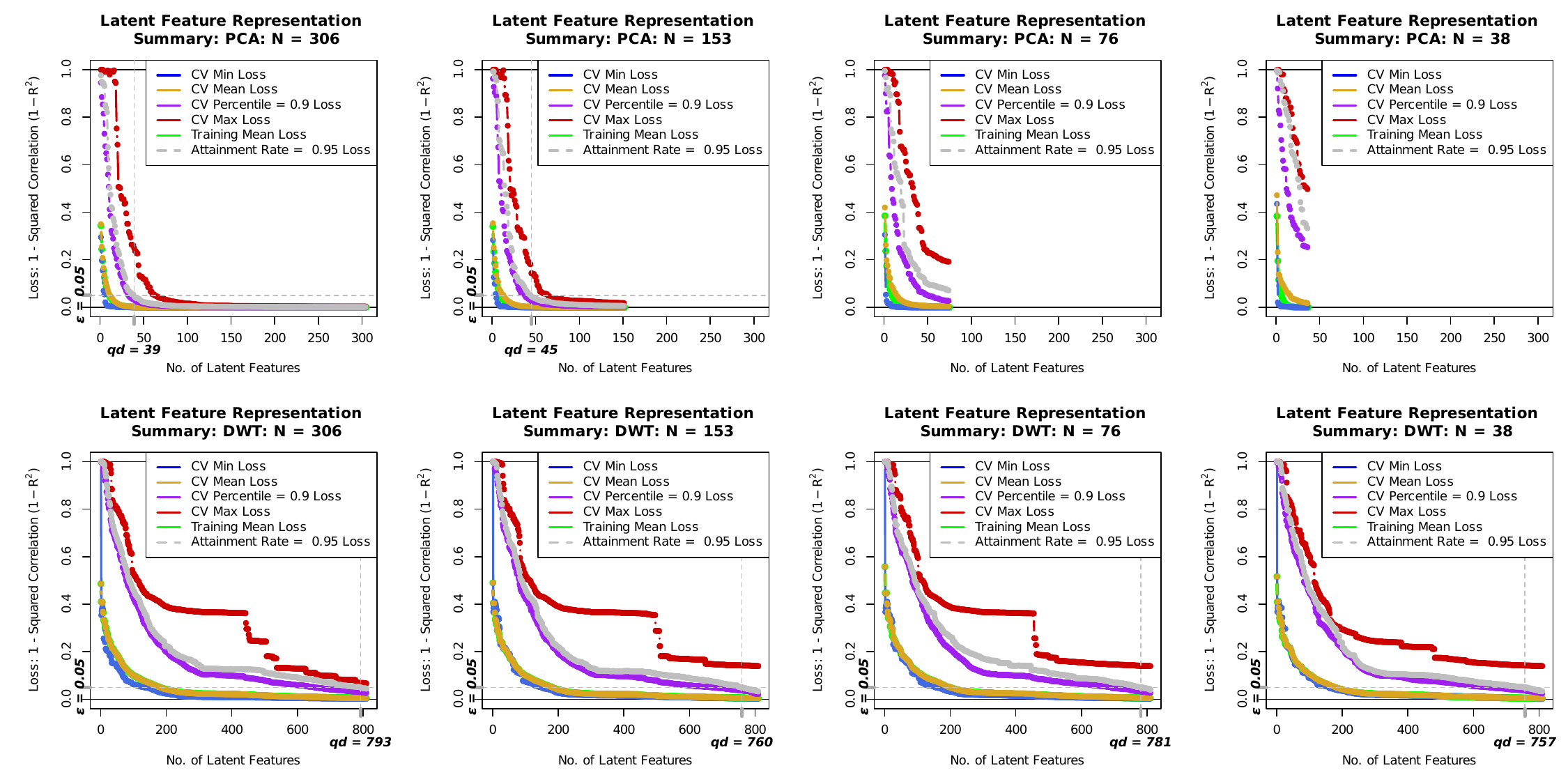}
    \caption{Results of the experiment to assess the effect of sample size on different latent feature representations. The Glaucoma data was used to create smaller datasets of size $N=153$, $N=76$ and $N=38$. \texttt{GLaRe()} was used to compare the representations provided by PCA (first row) and DWT (second row) as the sample size was decreased. Leave-one-out cross-validation was used in all cases.}
    \label{fig:eye-sample-size-results-results-01}
\end{figure}

Figure \ref{fig:eye-sample-size-results-results-01} displays the results of the experiment.
The PCA results are displayed in the top row and the DWT results are displayed in the bottom row.
As PCA can only estimate, at most, $\min(N-1, T)$ features and in this case $N<T$, we see that the maximum possible number of latent features changes as the sample size decreases.
In contrast, there is no restriction on the number of latent features for the DWT and we manually choose a maximum of $K=800$, which is sufficient to achieve the qualifying criterion (with $\epsilon=0.05$ and $1 - \alpha=0.95$) in all four cases.
The greater reliance of PCA on sample size is reflected in Figure \ref{fig:eye-sample-size-results-results-01} in several ways.
Firstly, the displayed quantiles of the cross-validated loss distribution, in particular the maximum, $0.95$ and $0.9$ quantiles (red, grey and purple lines) increase noticeably and systematically for PCA as the sample size is decreased, but they remain more stable for the DWT.
The dependence is also reflected by the separation between the training and validation mean losses (green vs. yellow lines) as sample size is decreased to $N=38$ for PCA but not the DWT, which demonstrates that PCA is unable to estimate a generalizable representation when the sample size is small, even if the training loss is satisfactory.
Finally, we can achieve the qualifying criterion using the DWT in all four cases (albeit with a large number of features), but we only achieve it for PCA with sample sizes $N=306$ and $N=153$.
The experiment is subject to sampling variability induced in the sub-sampling stage, so it is repeated using different random seeds in Appendix \ref{sec:additional-results}.

\section{Software Implementation (GLaRe)}\label{sec:software}

The CLaRe framework is implemented in our \proglang{R} package called \pkg{GLaRe}.
In this section, we provide an overview of the main components of the package, with additional details provided in Appendix \ref{sec:additional-outputs}.

The main function in the \pkg{GLaRe} package is \texttt{GLaRe()}, which computes and summarizes the cross-validated distribution of information loss and implements the CLaRe framework to compute the qualifying dimension for a choice of $1 - \alpha$ and $\epsilon$.
A sample call to the \texttt{GLaRe()} function is presented in Listing 1.

\begin{lstlisting}[language=R, caption={Example call to the \texttt{GLaRe()} function.}]
mnist_pca <- GLaRe(mat = X,
                   latent_dim_from = 1,
                   latent_dim_to = 400,
                   latent_dim_by = 20,
                   attainment_rate = 0.95,
                   tolerance_level = 0.05,
                   learn = "pca",
                   verbose = TRUE)
\end{lstlisting}
The \texttt{GLaRe()} function computes full cross-validated distribution of information losses for a latent feature representation method defined by \texttt{learn}, on dataset stored in the matrix \texttt{mat}, across a range of latent feature dimensions defined by the grid of equally-spaced values from \texttt{latent\_dim\_from} to \texttt{latent\_dim\_to} in increments of \texttt{latent\_dim\_by}.
It computes whether the method achieves the qualifying criterion defined by \texttt{tolerance\allowbreak\_level} $\epsilon$ and \texttt{attainment\_rate} $(1 - \alpha)$ for this range of latent feature dimensions and, if so, the learned latent feature representation at the qualifying dimension.
We describe the main features of the \texttt{GLaRe()} function in the following sections.

\subsection{Learning Functions}\label{sec:learning-functions}

The learning function is the main engine of \texttt{GlaRe()}.
The learning function takes as arguments a data matrix $\mathbf{X}$ and latent feature dimension $K$, and learns the encoding and decoding transformation functions $f_K$ and $g_K$.
For example, the learning function in PCA computes the first $K$ eigenvectors of the empirical covariance matrix of $\mathbf{X}$, and $f_K$ and $g_K$ comprise matrix multiplication by the first $K$ eigenvectors.
In contrast, for an autoencoder, $f_K$ and $g_K$ are general functions that map to and from a $K$-dimensional latent space and are parametrized by flexible neural networks.
In \texttt{GLaRe()}, the learning function is defined by the \texttt{learn} argument. 
We provide three built-in learning functions that can be used, as well as the option to specify a bespoke, user-defined learning function:
\begin{enumerate}
    \item Setting \texttt{learn = "pca"} specifies a PCA representation. The eigenvectors are computed by the Singular Value Decomposition (SVD) algorithm. In cases where $N < T$ (e.g., the Glaucoma or Gels data) the latent feature dimension for PCA can be, at most, $N-1$. Hence the maximum latent feature dimension \texttt{latent\_dim\_to} is set to a default of $\min(N-1, T)$ when using PCA.
    \item Setting \texttt{learn = "dwt"} (or \texttt{learn = "dwt.2d"} for data on a 2-dimensional domain) specifies a thresholded wavelet representation. For encoding, the DWT is applied to $\mathbf{X}$ and the most important $K$ latent features (i.e., wavelet coefficients) are learned from the data and are retained. The decoding function then applies the inverse DWT to the retained features. Our implementation uses the \pkg{wavselim} \proglang{R} package \parencite{whitcher_waveslim_2024} which uses the Daubechies orthonormal compactly supported wavelet of length $8$ \parencite{daubechies_ten_1992}, least asymmetric family and uses periodic boundary conditions. Additional details are described in Appendix \ref{sec:wavelet-thresholding-algorithm}.
    \item Setting \texttt{learn = "ae"} specifies an autoencoder representation. We implement the autoencoder using the \pkg{keras} \proglang{R} package \parencite{kalinowski_keras_2024}. The encoder and decoder functions are parametrized by neural networks with a single hidden layer (defaulting to size $600$) and a rectified linear unit (ReLU) activation function.
    A linear activation function is used to map the hidden layer of the encoder to the latent space, and either a linear and sigmoid (default) activation function can be used to map from the hidden layer of the decoder back to the data space.
    By default, the autoencoder is trained for $100$ epochs using the ADAM stochastic gradient descent algorithm \parencite{kingma_adam_2017} to minimize either the mean squared error (default) or binary cross-entropy loss functions using a mini-batch size of $16$.
    \item Setting \texttt{learn = "user"} allows the user to specify their own latent feature representation method. With this setting, the user must supply the learning function for their method, that takes the data matrix $\mathbf{X}$ and the latent feature dimension $K$ as inputs and returns a list with two elements: functions named \texttt{Encode} and \texttt{Decode} implementing the learned encoding and decoding transformation functions $f_K$ and $g_K$.
\end{enumerate}

\subsection{Squared Correlation Loss}\label{sec:loss-functions}

In principle, any loss function that satisfies the properties outlined in Section \ref{sec:materials-and-methods} could be used with the \texttt{GLaRe()} function and the package has been structured such that different loss functions can be used in future iterations.
The current implementation uses the complement of the squared correlation
$$
1- \rho^2 \left\{X_i (\mathbf{t}), \widehat{X}^{(K)}_{i} (\mathbf{t}) \right\} =
1 - \frac{\left[ \mathlarger{\sum}_{t = 1}^T{\bigg\{X_i (t) - \overline{X}_i \bigg\} \bigg\{ \widehat{X}_i^{(K)} (t) - \overline{\widehat{X}}_i^{(K)} \bigg\}} \right]^2}{\mathlarger{\sum}_{t = 1}^T \bigg\{X_i (t) - \overline{X}_i \bigg\}^2 \mathlarger{\sum}_{t = 1}^T \bigg\{ \widehat{X}_i^{(K)} (t) - \overline{\widehat{X}}_i^{(K)} \bigg\} ^2},
$$
where
$$
\overline{X}_i = \frac{1}{N} \sum_{t=1}^T X_i (t) \quad \text{and} \quad \overline{\widehat{X}}_i^{(K)} = \frac{1}{N} \sum_{t=1}^T \widehat{X}_i^{(K)} (t).
$$
The criterion $1- \rho^2$ is a relative measure, bounded between $0$ and $1$: $1- \rho^2 = 0$ indicates losslessness and $1- \rho^2 = 1$ indicates that no information is preserved by the latent feature representation method.
Employing a relative measure ensures that choices for the tolerance level $\epsilon$ are comparable across datasets that differ in scale and dimensionality.
One disadvantage of this measure is that it is undefined when the predicted value is constant, we set $1 - \rho^2 = 1$ in our software when this happens. 
In Appendix \ref{sec:squared-correlation}, we highlight connections between squared correlation and the Predicted Residual Sum of Squares (PRESS) in the context of PCA.

\subsection{Cross-Validation Algorithm}

Given the learning function (Section \ref{sec:learning-functions}), the loss function (Section \ref{sec:loss-functions}), the data matrix $\mathbf{X}$ and a grid of values for $K$, \texttt{GLaRe()} implements a $k$-fold cross validation algorithm to estimate the individual generalization errors (Algorithm 1).
In principle, it is possible to employ leave-one-out cross-validation by setting the number of folds to the number of observations $N$.
For leave-one-out cross-validation, the division into folds is systematic and does not induce any variability.
However, in most real-world applications we employ $5$ or $10$-fold cross-validation as an approximation to leave-one-out due to computational considerations.
In these cases, variability is induced by splitting of the data into folds (i.e., the data rows are randomly shuffled and then divided into folds), so we recommend fixing the random seed before running \texttt{GLaRe()} so that results are reproducible.
It can also be useful to re-run \texttt{GLaRe()} using different seeds to assess the sensitivity of results to the random split and to decide whether to employ a larger number of folds or even leave-one-out cross-validation.

\subsection{Software Outputs}

The \texttt{GLaRe()} function computes and returns the cross-validated information loss for each individual observation in the dataset.
It provides the custom graphic and the heatmap shown in Figure \ref{fig:viz-options} to summarize this distribution of information losses graphically.
If the qualifying criterion is met within the range of latent feature dimensions, the final model is fit to the full dataset at the qualifying dimension and the learned encoding and decoding transformation functions are returned as \proglang{R} functions called \texttt{Encode()} and \texttt{Decode()}.
Further details on the graphical outputs from GLaRe are described in Appendix \ref{sec:additional-outputs}.

\section{Discussion}\label{sec:discussion}

We have introduced CLaRe, an evaluation framework for assessing and comparing latent feature representation methods with a focus on compactness and near-losslessness.
A distinguishing feature of CLaRe is its focus on generalized quantiles of the distribution of generalization errors rather than aggregated metrics (e.g., average or total information loss) that reflect the distribution's central tendency.
Our framework uses cross-validation to estimate the full distribution of generalization errors and proposes to choose a representation such that a tolerance level of generalization error is met for generalized quantiles of this distribution (e.g., worst case or $95$th percentile) while ensuring that the representation is as compact as possible.
Thus, CLaRe enables the selection of a compact, near-lossless latent feature representations that ensures statistical modeling in the latent space can accurately reflect the underlying mechanisms of the true data-generating process.

One of CLaRe's strengths is that it facilitates comparisons between methods, allowing comparisons between traditional tools such as PCA and modern approaches such as autoencoders. 
Through case studies on three datasets—Glaucoma, Proteomic Gels, and MNIST, we have demonstrated how CLaRe can guide the selection of the most suitable latent feature representation based on dataset characteristics.
The results from these case studies reinforce the importance of such context-specific evaluation, as the preferred representation method varied across datasets. 
For instance, the MNIST dataset, with its large sample size relative to feature dimension, benefited from the flexibility of the non-linear AE representation. 
In contrast, for the Proteomic Gels dataset, which is characterized by a small sample size relative to its high-dimensional features, the fixed DWT representation was preferred to the more flexible PCA and AE representations.
We performed an experiment by manually decreasing the sample size of the Glaucoma dataset and comparing the PCA and DWT representations which further highlighted the trade-off between flexible methods and sample size.
These case studies emphasize the role of dataset characteristics, such as sample size, dimensionality and variance structure, in determining the most appropriate latent feature representation. 
CLaRe is a valuable framework to compare methods under a consistent set of criteria in such contexts.

We have also described and documented our accompanying \proglang{R} package, GLaRe, which implements the CLaRe framework and provides intuitive graphical summaries for the user.
GLaRe provides a flexible implementation of the framework where the user can specify the criteria (e.g., tolerance level, attainment rate).
It provides built-in implementations of three latent feature representation methods -- PCA, DWT and autoencoder -- but it also allows the user to easily specify a latent feature representation method of their own. The package is publicly available and can be employed by practitioners in any analysis that relies on latent feature representation methods.

Some limitations and future directions of this work are as follows.
Our framework focuses on compactness and near-losslessness, which are two of the most important properties of a latent feature representation.
However, other properties might also need to be considered when selecting among representations, e.g., distribution and dependence structure of features in the latent space, interpretability of the latent features, computational time, and effort.
In some situations, e.g., data measured with white noise errors, attaining the (near-)losslessness property might not be possible.
However, in these situations, it is still useful to quantify information loss at an individual observation level so that it can acknowledged when presenting the results of subsequent modeling in the latent space.
The current framework should also be extended to handle dependent (e.g., multilevel, longitudinal, temporal/ spatial) data by including structured variants of cross-validation for dependent datasets
\parencite{bergmeir_note_2018, hornung_evaluating_2023, roberts_cross-validation_2017}.
In our case studies, we used standard versions of PCA, DWT and AE to facilitate general and straightforward comparisons but future work could consider specialized implementations (e.g., smoothed functional PCA or convolutional autoencoders).
While our framework immediately extends to more general non-Euclidean data objects, e.g., shapes \parencite{srivastava_shape_2011}, trees \parencite{wang_object_2007}, probability distributions \parencite{petersen_functional_2016, yang_quantile_2020} and correlation/ covariance matrices \parencite{desai_connectivity_2023}, specialized transformations that encourage a Euclidean structure in the latent space and preserve essential properties when mapping back to the data space would need to be considered.
\section*{Supporting Information}

Appendices \ref{sec:additional-data}-\ref{sec:additional-results} contain additional details of the analysis.
\proglang{R} code scripts and data to reproduce the analysis are available at \proglang{GitHub}\footnote{ \url{https://github.com/edwardgunning/MANUSCRIPT-CLaRe}.}.
\section*{Acknowledgements}
Emma Zohner and Edward Gunning are co-first authors of this work.
We are grateful to Martin Das (Sr. Software and Technical Specialist) for his computational assistance.
This work was partially supported by the following grants: CA-244845 and CA-178744.
\clearpage

\printbibliography
\clearpage

\appendix
\section{Additional Data Example: The \texttt{phoneme} Dataset}\label{sec:additional-data}

\begin{figure}[h]
    \centering
    \includegraphics[width=0.75\linewidth]{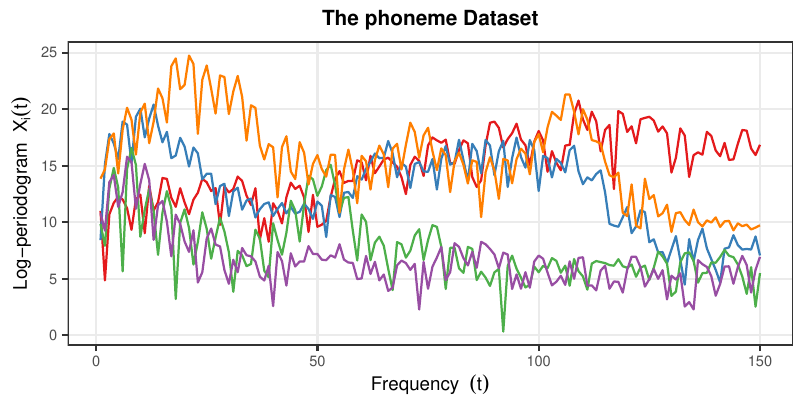}
    \caption{A random sample of $8$ observations from the \texttt{phoneme} dataset \parencite{hastie_elements_2009, ferraty_nonparametric_2006}.}
    \label{fig:phoneme}
\end{figure}

The \texttt{phoneme} dataset, from the book \emph{Nonparametric Functional Data Analysis: Theory and Practice} by \textcite{ferraty_nonparametric_2006}\footnote{Data available at \url{https://www.math.univ-toulouse.fr/~ferraty/SOFTWARES/NPFDA/}}, is a dataset from the field of speech recognition analysis.
It emanates from an original dataset used in the book \emph{The Elements of Statistical Learning} by \textcite{hastie_elements_2009}\footnote{Data available at \url{https://hastie.su.domains/ElemStatLearn/}}.
The data contains observations of an audio signal that is transformed to the log-periodogram scale at a range of frequencies.
\textcite{ferraty_nonparametric_2006} provide $N=2000$ observations of the signal discretized onto a grid of $T=150$ equidistant frequencies, so that $\mathbf{X}$ is a $2000\times 150$ matrix containing the signals in its rows.
Figure \ref{fig:phoneme} displays a random sample of $8$ observations from the dataset.

\begin{figure}
    \centering
    \includegraphics[width=1\linewidth]{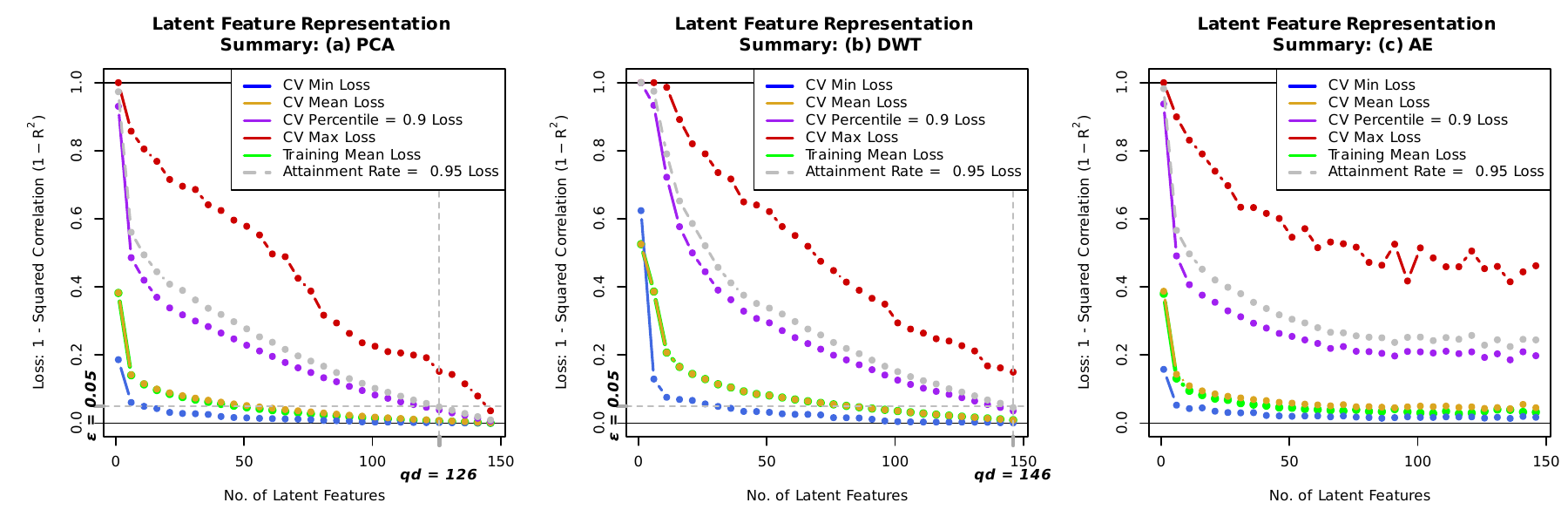}
    \caption{Summary \texttt{GLaRe()} plot for the \texttt{phoneme} data. A grid of equally-spaced values from $1$ to $150$ in increments of $5$ was used for the latent feature dimensions.}
    \label{fig:phoneme-results}
\end{figure}

Figure \ref{fig:phoneme-results} displays the results of applying our CLaRe framework to select among PCA, DWT and AE representations for the phenome dataset. A grid of equally-spaced values from $1$ to $150$ in increments of $5$ was used for the latent feature dimensions.
The qualifying criterion was achieved for PCA and DWT but not for the AE. The qualifying dimensions for PCA and DWT were $qd=126$ and $qd = 146$, respectively. Hence, PCA was the favored latent feature representation method for this dataset.

\section{Additional Details on Software Outputs and Functionality}\label{sec:additional-outputs}

\begin{figure}
    \centering
    \includegraphics[width=0.5\linewidth]{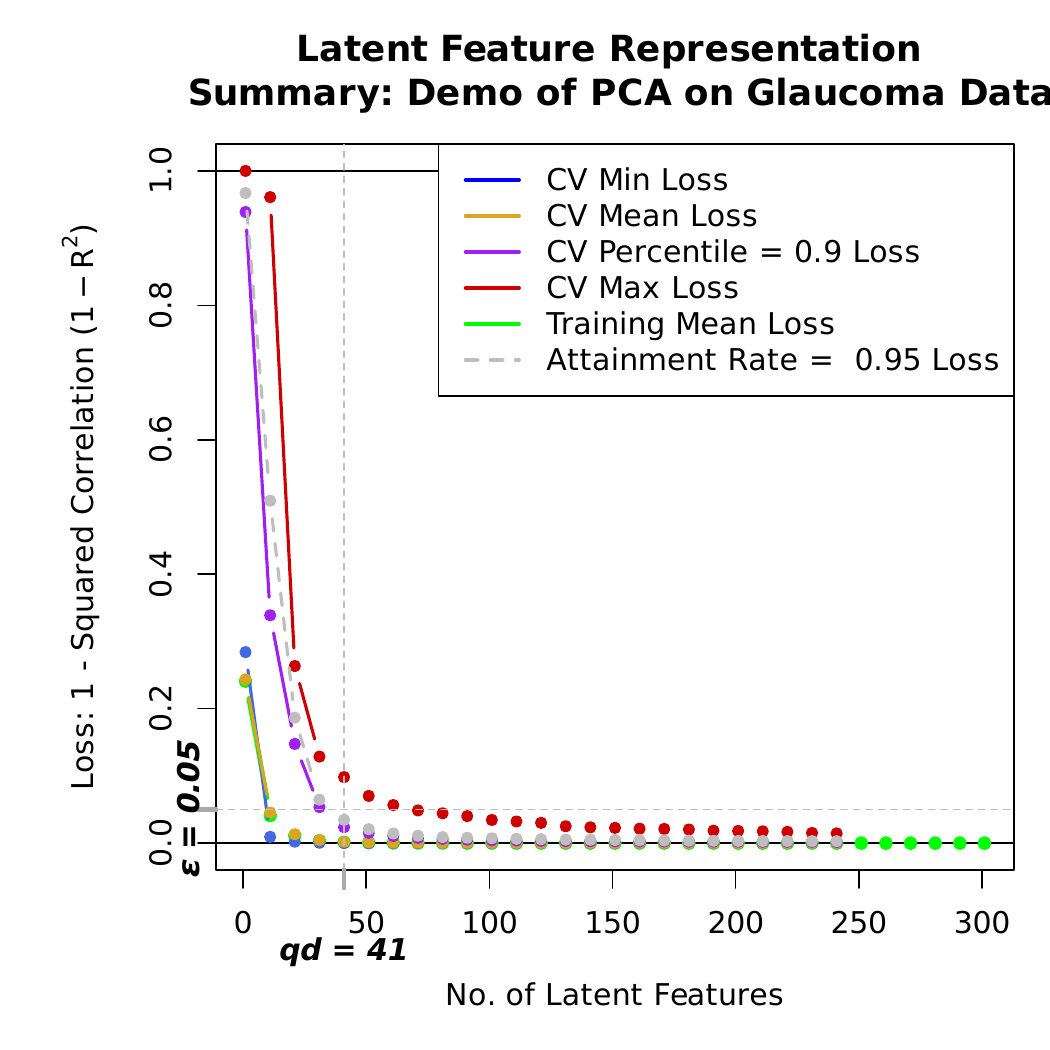}
    \caption{The summary plot produced by \texttt{GlaRe()}, demonstrated on the Glaucoma dataset with PCA.}
    \label{fig:glare-anatomy-plot}
\end{figure}

The \texttt{GLaRe()} summary plot is the default graphic returned by the \proglang{GLaRe()} function (Figure \ref{fig:glare-anatomy-plot}).
The overall, or average, cross-validated loss is displayed in yellow, with the analogous loss computed on the training data shown in green for comparison.
Then, different quantiles of the distribution of individual cross-validated losses are displayed to summarize the full distribution: the minimum and maximum are shown in blue and red, respectively, a user-specified quantile of the distribution (set by the \texttt{cvqlines} argument) is displayed in purple and the quantile of the distribution being used as the attainment rate $1 - \alpha$ (defaulting to 0.95, i.e., the $95$th percentile) is displayed in light gray.
The corresponding value of the tolerance level $\epsilon$ is overlaid as gray dashed horizontal line, and hence the latent feature dimension (i.e., location on the $x$-axis) at which the two gray lines meet corresponds to the qualifying dimension.
The tolerance level $\epsilon$ and the qualifying dimension ($qd$) are marked in bold and italic typeface on the $y$ and $x$ axes, respectively.
The \texttt{GLaRe()} function also returns a heatmap as an alternative summary of the individual cross-validated loss distribution, and the package also contains wrapper functions to display a dot-plot of the full distribution of cross-validated losses, the ratio of training to cross-validated losses and the reconstructions of individual observations at the qualifying dimension. 

\begin{figure}
    \centering
    \includegraphics[width=0.5\linewidth]{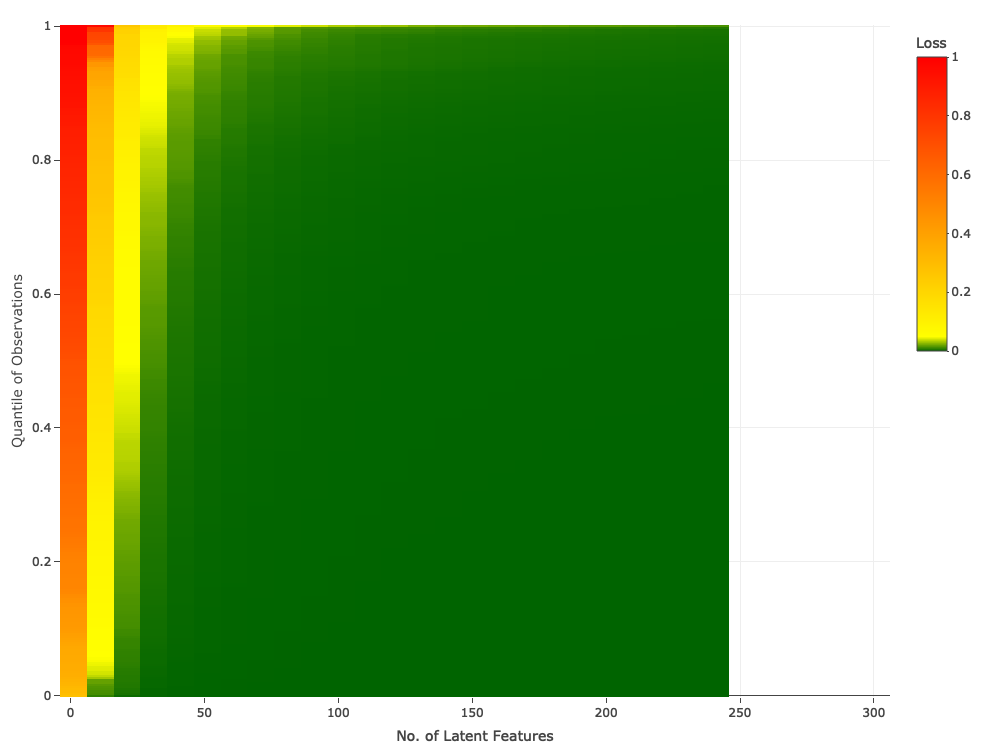}
    \caption{The heatmap returned by \texttt{GLaRe()} used to summarize the full distribution of generalization errors (i.e., cross-validated estimates of information loss), demonstrated on the Glaucoma dataset with PCA. The latent feature dimension is represented on the $x$-axis, the corresponding quantile of the generalization error distribution at that feature dimension is shown on the $y$-axis and the color represents the value of the generalization error at that feature dimension and quantile.}
    \label{fig:eye-heatmap}
\end{figure}

The \texttt{GLaRe()} function also returns a heatmap to display the full distribution of generalization errors (Figure \ref{fig:eye-heatmap}).
It is obtained by re-ordering the $N$ values within each column of the matrix of cross-validated information losses.
The latent feature dimension is represented on the $x$-axis, the corresponding quantile of the generalization error distribution at that feature dimension (i.e., column) is shown on the $y$-axis and the color represents the value of the generalization error at that feature dimension and quantile.

\begin{figure}
    \centering
    \includegraphics[width=1\linewidth]{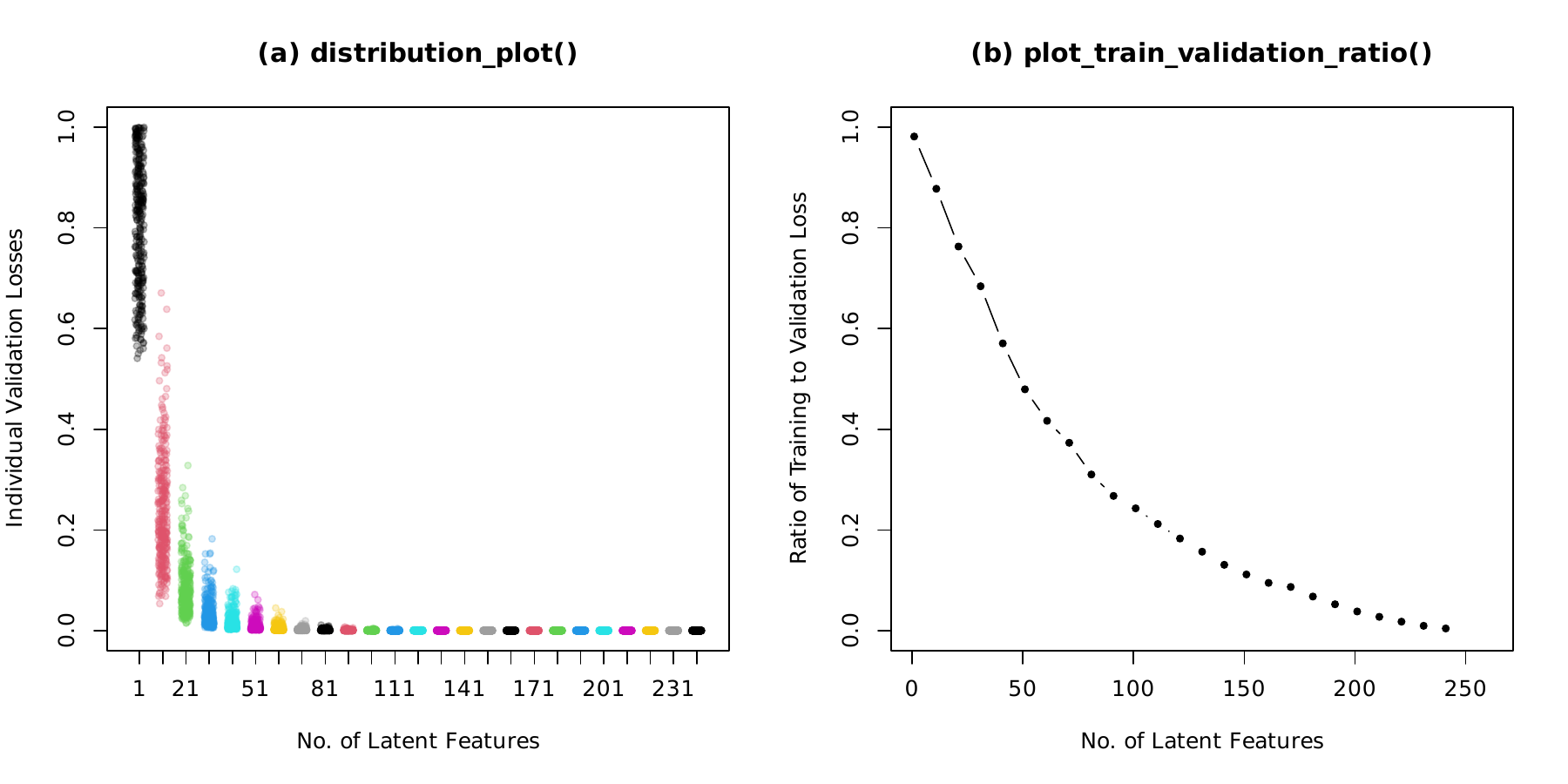}
    \caption{Additional wrapper functions that produce summary plots of \texttt{GLaRe()} outputs. \textbf{(a)} \texttt{distribution\_plot()} produces a dot-plot of the individual cross-validated information loss distribution. \textbf{(b)} \texttt{plot\_train\_validation\_ratio()} produces a point and line plot of the ratio of the total training and validation losses.
    Both plots are demonstrated on the Glaucoma dataset with a PCA representation from Section \ref{sec:software}.}
    \label{fig:additional-plots-01}
\end{figure}

The \pkg{GLaRe} package also contains wrapper functions that plot alternative summaries of the cross-validated distribution of information losses; their outputs are displayed in Figure \ref{fig:additional-plots-01}.
The function \texttt{distribution\_plot()} produces a dot-plot of the individual cross-validated information loss distribution, where each point represents an individual value and the points are colored according to the latent feature dimension $K$ (Figure \ref{fig:additional-plots-01} \textbf{(a)}).

\begin{figure}
    \centering
    \includegraphics[width=0.75\linewidth]{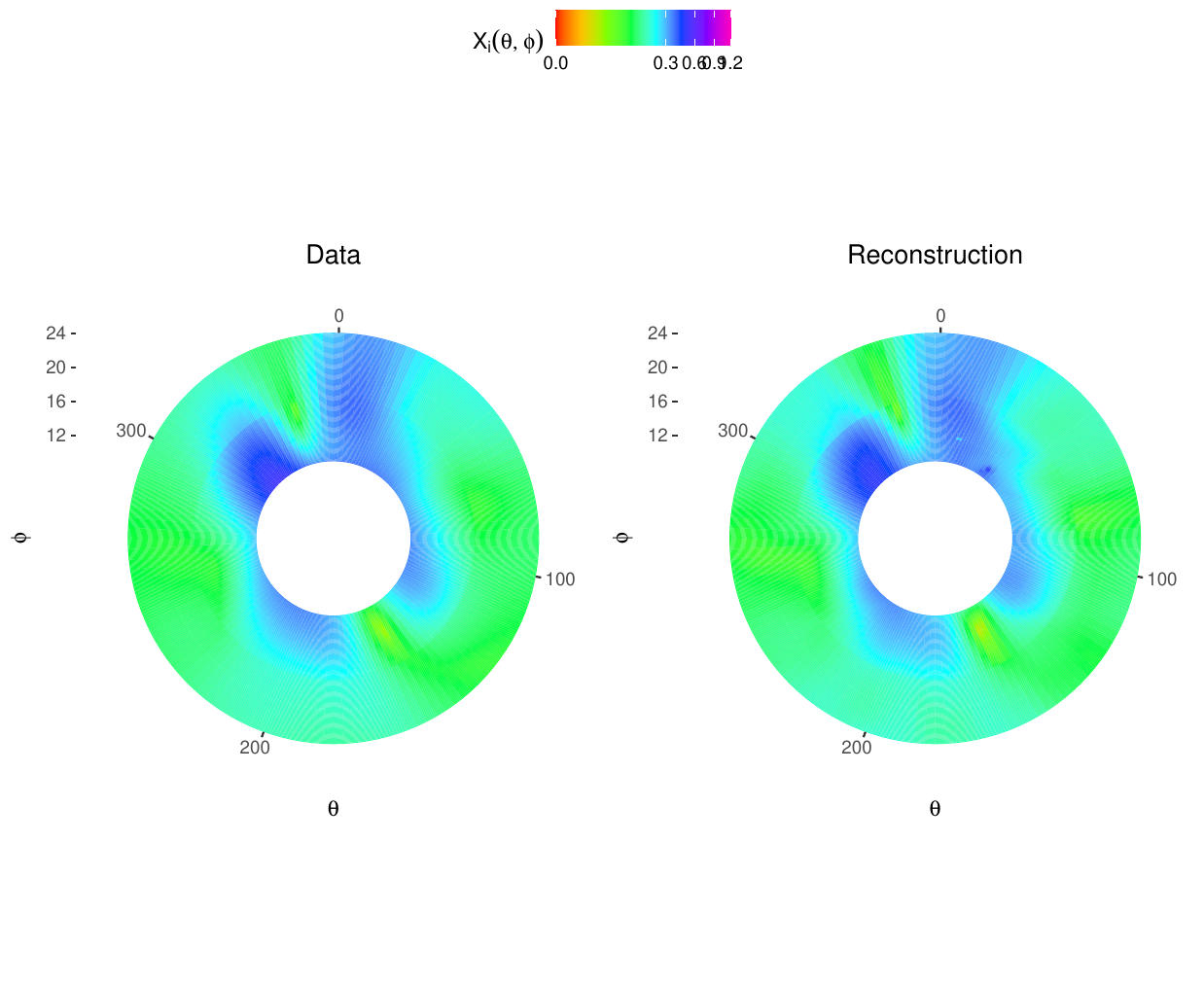}
    \caption{A single observation from the Glaucoma data (left) and its reconstruction (right) using the final model fit from PCA at the qualifying dimension $qd = 51$. The figure was generated using the \texttt{plot\_eye\_reconstruction()} function from the \pkg{GLaRe} package. A cubed-root transformation is applied for visualization.}
    \label{fig:eye-reconstruction}
\end{figure}

\begin{figure}
    \centering
    \includegraphics[width=0.75\linewidth]{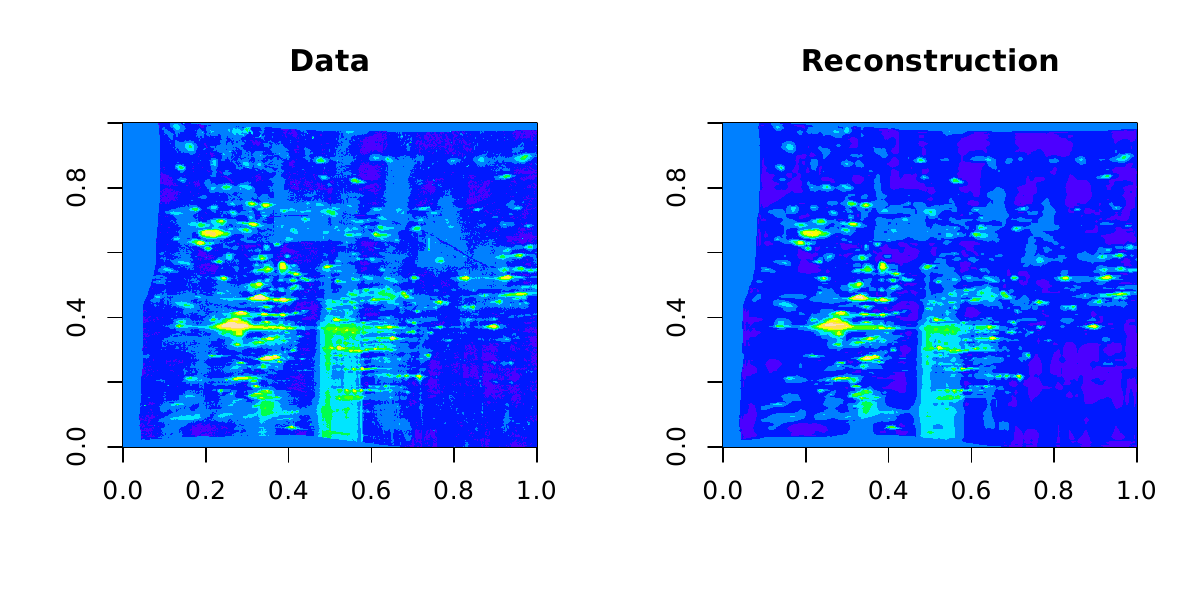}
    \caption{A single observation from the Proteomic Gels data (left) and its reconstruction (right) using the final model fit from DWT at the qualifying dimension $qd = 7801$. The figure was generated using the \texttt{plot\_gels\_reconstruction()} function from the \pkg{GLaRe} package.}
    \label{fig:gels-reconstruction}
\end{figure}

\begin{figure}
    \centering
    \includegraphics[width=0.75\linewidth]{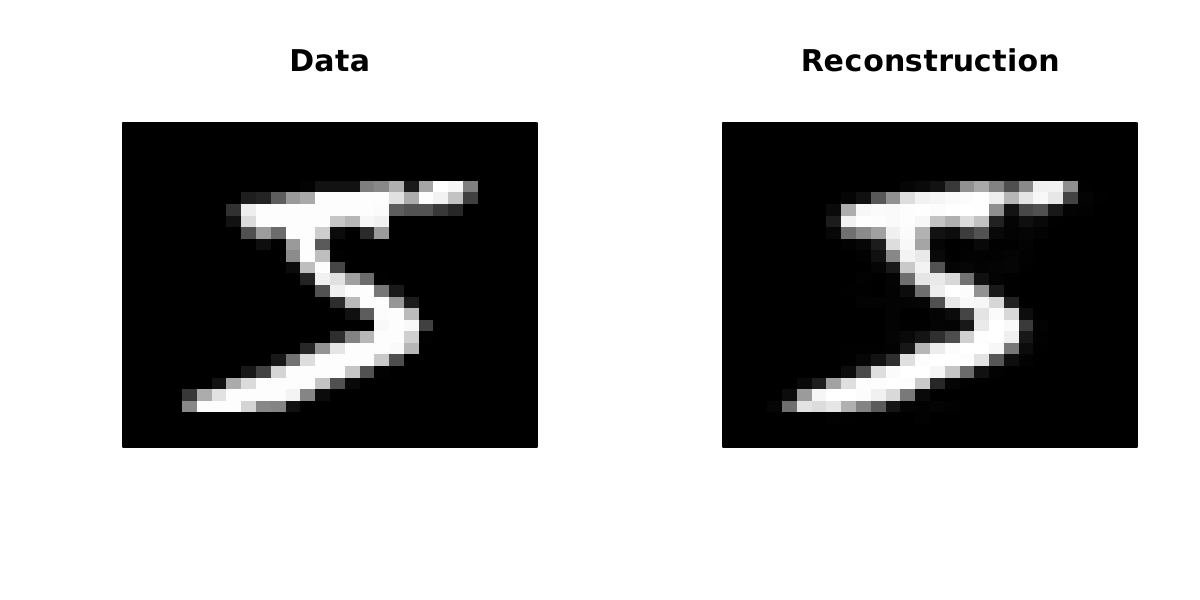}
    \caption{A single observation from the MNIST Digits data (left) and its reconstruction (right) using the final model fit from AE at the qualifying dimension $qd = 101$. The figure was generated using the \texttt{plot\_mnist\_reconstruction()} function from the \pkg{GLaRe} package.}
    \label{fig:mnist-reconstruction}
\end{figure}

\begin{figure}
    \centering
    \includegraphics[width=0.75\linewidth]{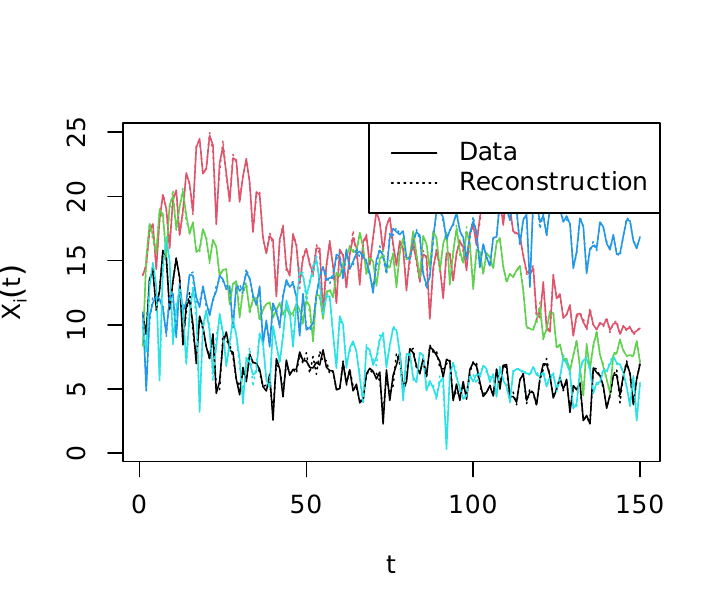}
    \caption{Reconstructions of $8$ observations from the \texttt{phenome} dataset (dotted lines) overlaid on the original observations (solid lines). The reconstructions were computed from the final model fit of PCA at the qualifying dimension $qd=126$. The figure was generated using the \texttt{plot\_1D\_reconstruction()} function from the \pkg{GLaRe} package.}
    \label{fig:phoneme-reconstruction}
\end{figure}

When the qualifying criterion is met, the CLaRe framework fits the final model on the full dataset at the qualifying dimension.
The \pkg{GLaRe} package contains functions to visually inspect the reconstruction of individual observations from the final model.
Due to the non-standard structure of the Glaucoma, Proteomic Gels and MNIST data, we have written custom functions to display side-by-side plots of the data observation and its reconstruction (Figures \ref{fig:eye-reconstruction} -- \ref{fig:mnist-reconstruction}).
For data objects which are $1$-dimensional signals, we provide a general function called \texttt{plot\_1D\_reconstruction()} that displays the original signal as a solid line and overlays its reconstruction as a dotted line (Figure \ref{fig:phoneme-reconstruction}).
Unlike the specialized functions that plot side-by-side plots, this function can display the reconstruction of more than one observation simultaneously.

\section{Wavelet Thresholding Algorithm}\label{sec:wavelet-thresholding-algorithm}

We use the Discrete Wavelet Transform (DWT) algorithm implementation in the \texttt{dwt()} function from the \pkg{wavselim} \proglang{R} package \parencite{whitcher_waveslim_2024}.
The thresholding approach we present below is described in full in Section 3.5 of \textcite{morris_automated_2011}. Here we demonstrate it on the \texttt{DTI} dataset from the \pkg{refund} \proglang{R} package \parencite{goldsmith_refund_2020}.
As our main focus is on the technical steps of the wavelet decomposition, we describe the selection of the coefficients on the same sample of data that we want to reconstruct. However, in practice, Steps 3 and 4 would only be performed on the training data and then validation data used to assess generalization error using a chosen truncation.

\begin{steps}
  \item \underline{\textbf{Pad the Data to Dyadic Length}}: The DWT can only be applied to vectors of dyadic length, i.e., a power of $2$. In most cases, the we work with the $N \times T$ data matrix $\mathbf{X}$ where $T$ is not a power of $2$ (i.e., $\log_2(T)$ is not an integer). If this is the case, we define $\log_2(T_{pad})$ as the smallest integer greater than $\log_2(T)$. We then add $\lceil (T_{pad} - T)/2 \rceil$ columns of $0$'s to the left and $\lfloor (T_{pad} - T)/2 \rfloor$ columns of $0$'s to the right side of $\mathbf{X}$, so that the resulting matrix $\mathbf{X}_{pad}$ has dimensions $N \times T_{pad}$ (Figure \ref{fig:DTI-padded}).
  \begin{figure}[H]
      \centering
      \includegraphics[width=0.75\linewidth]{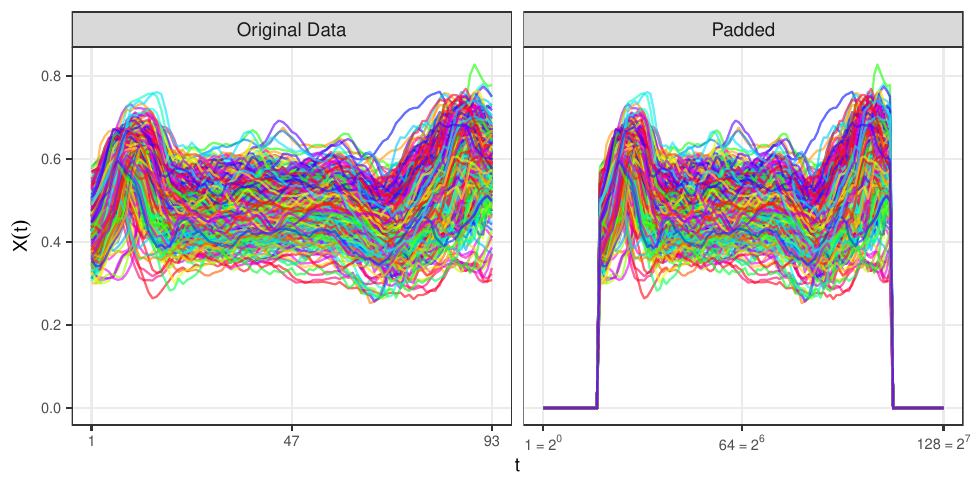}
      \caption{Padding the \texttt{DTI} data to transform it from length $T = 93$ to $T_{pad} = 128 = 2^7$.}
      \label{fig:DTI-padded}
  \end{figure}
  \item \underline{\textbf{Apply the DWT to Each Row}}: We then apply the DWT to each row of $\mathbf{X}_{pad}$, which transforms each vector from $T_{pad}$ measurements of a time series (or signal) to $T_{pad}$ wavelet coefficients.
  By default, the \texttt{waveslim::dwt()} function employs ``the Daubechies orthonormal compactly supported wavelet of length L=8 \parencite{daubechies_ten_1992}, least asymmetric family" as the wavelet filter, with periodic assumptions for the signal beyond the boundaries \parencite[][p.7]{whitcher_waveslim_2024}.
  We add store the wavelet coefficients for each row in the rows of the $N \times T_{pad}$ matrix $\mathbf{X}^*$.
  When we have expanded the original signal by padding in Step 1, we can expect a number of the $T_{pad}$ wavelet coefficients to be $0$, however this number is likely to be less than $T_{pad} - T$.
  \item \underline{\textbf{Compute the Relative Energy Matrix}}: For each row of $\mathbf{X}^*$, we have the vector of wavelet coefficients $\mathbf{X}^*_{i\cdot} = (X^*_{i1}, \dots,X^*_{iT_{pad}})^\top$. We denote the \emph{Total Energy} for the $i$th observation as the sum of its squared wavelet coefficients
  $$\text{Total Energy}_i = \sum_{k=1}^{T_{pad}}X^{*2}_{ik}.$$ Next, we define the \emph{Cumulative Relative Energy} for the $i$th observation and wavelet coefficient $k$ as 
  $$
  \text{Relative Energy}_{ik} = \frac{\sum_{\{k: \lvert X^*_{ik'}\rvert  \geq \lvert X^*_{ik}\rvert \}}X^{*2}_{ik}}{\text{Total Energy}_i}.
  $$
  This quantity represents the proportion of the total energy that is explained by the $k$th wavelet coefficient and all coefficients greater in absolute value than it. Hence, smaller values indicate this coefficient is important and values closer to $1$ indicate less importance (i.e., a value of $1$ indicates that all of the energy has been explained before this coefficient).
  Normalising by the total energy is important because we summarise this quantity across all $i$ as a measure of importance in the next step, and the normalisation ensures that it the importance is not obscured by the total energy of an individual signal. We let $\textbf{En}^*$ represent the total energy matrix which contains $\text{Relative Energy}_{ik}$ in its $i$th row and $k$th column.
  \item \underline{\textbf{Compute the Relative Energy Scree}}: To summarise the overall importance of each of the wavelet coefficients we average each column of the matrix $\textbf{En}^*$. We obtain the $T_{pad}$-dimensional \emph{Scree} vector, that has the $k$th entry
  $$
  \text{Scree}_k = \frac{1}{N} \sum_{i=1}^N \text{Relative Energy}_{ik}.
  $$
  As with the individual relative energy matrix, coefficients with a lower average value are of greater importance while larger average values (closer to $1$) indicate less importance.
  \item \underline{\textbf{Hard Thresholding Based on the Relative Energy Scree}}: For a given $K < T_{pad}$, we threshold the wavelet coefficient matrix $\mathbf{X}^*$ based on the relative energy scree. That is, we retain the $K$ columns of $\mathbf{X}^*$ that have the smallest values of $\text{Scree}_k$ and set the remaining columns to $0$. We denote the thresholded version of $\mathbf{X}^*$ by $\widehat{\mathbf{X}}^{*(K)}$.
  \item \underline{\textbf{Apply IDWT to Each Row of the Thresholded Coefficient Matrix}}: To transform back to the data space, we apply the inverse DWT (IDWT) to each row of $\widehat{\mathbf{X}}^{*(K)}$. This will give use the reconstructed matrix $N \times T_{pad}$
  $$
  \widehat{\mathbf{X}}^{(K)}_{pad} = \text{IDWT}(\widehat{\mathbf{X}}^{*(K)}).
  $$
  To obtain a representation of the original signal length we simply discard the first $\lceil (T_{pad} - T)/2 \rceil$ columns and the last $\lfloor (T_{pad} - T)/2 \rfloor$ columns to give the matrix $\widehat{\mathbf{X}}^{(K)}$. Figure \ref{fig:DTI-recon} displays the reconstruction of the \texttt{DTI} data using differing values of $K$, alongside the original data.
  \begin{figure}[H]
      \centering
      \includegraphics[width=0.8\linewidth]{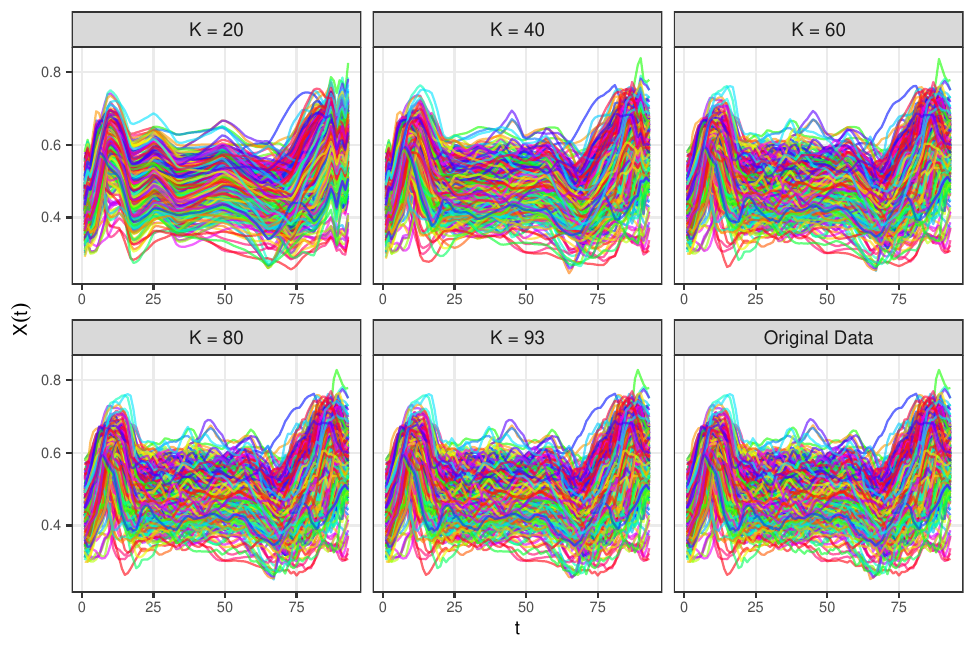}
      \caption{Thresholded wavelet representations of the \texttt{DTI} data with differing values of $K$, alongside the original data (bottom right).}
      \label{fig:DTI-recon}
  \end{figure}
\end{steps}

\section{Additional Details on Squared Correlation Loss} \label{sec:squared-correlation}

\subsection{Complement of the Predicted Correlation Squared ($1-\rho^2$)}

Our loss measure is the complement of the squared correlation among the observed data $X_i (\mathbf{t})$ and its predicted value $\widehat{X}^{(K)}_{i} (\mathbf{t})$:
$$
1- \rho^2 \left\{X_i (\mathbf{t}), \widehat{X}^{(K)}_{i} (\mathbf{t}) \right\} =
1 - \frac{\left[ \mathlarger{\sum}_{t = 1}^T{\bigg\{X_i (t) - \overline{X}_i \bigg\} \bigg\{ \widehat{X}_i^{(K)} (t) - \overline{\widehat{X}}_i^{(K)} \bigg\}} \right]^2}{\mathlarger{\sum}_{t = 1}^T \bigg\{X_i (t) - \overline{X}_i \bigg\}^2 \mathlarger{\sum}_{t = 1}^T \bigg\{ \widehat{X}_i^{(K)} (t) - \overline{\widehat{X}}_i^{(K)} \bigg\} ^2},
$$
where
$$
\overline{X}_i = \frac{1}{N} \sum_{t=1}^T X_i (t) \quad \text{and} \quad \overline{\widehat{X}}_i^{(K)} = \frac{1}{N} \sum_{t=1}^T \widehat{X}_i^{(K)} (t).
$$

\subsection{Predicted Residual Sum of Squares (PRESS)}

The predicted residual sum of squares (PRESS) statistics measures the discrepancy, in terms of total squared error, between the observed data $X_i (\mathbf{t})$ and its predicted value $\widehat{X}^{(K)}_{i} (\mathbf{t})$. 
The PRESS statistic is defined as the squared Euclidean distance between the observed data and its predicted value:
\begin{align*}
    \text{PRESS}\left\{X_i (\mathbf{t}), \widehat{X}^{(K)}_{i} (\mathbf{t}) \right\} 
&= 
\sum_{t = 1}^T \left\{ X_{i}(t) - \widehat{X}^{(K)}_{i} (t)\right\}^2 \\
&=
\bigg\| X_i (\mathbf{t}) - \widehat{X}^{(K)}_{i} (\mathbf{t}) \bigg\|^2\\
&=
\left\{X_i (\mathbf{t}) - \widehat{X}^{(K)}_{i} (\mathbf{t})\right\}^\top \left\{X_i (\mathbf{t}) - \widehat{X}^{(K)}_{i} (\mathbf{t})\right\}.
\end{align*}
While we have defined the PRESS statistic for individual observations, given our focus on individual information loss values, the total PRESS statistic summed over all observations is typically used to summarize the information loss in a PCA representation \parencite{bro_cross-validation_2008}.

\section*{Proof: Relationship between PRESS and $1 -\rho^2$ for PCA-based Projections}

When PCA is employed, $\widehat{X}^{(K)}_{i}(\mathbf{t})$ represents the projection of $X_{i}(\mathbf{t})$ onto \( K \)-dimensional subspace spanned by the first $K$ eigenvectors from PCA. That is
$$
 \widehat{X}^{(K)}_{i} (\mathbf{t}) = \underbrace{\boldsymbol{\Phi}_K}_{\text{Eigenvectors}} \underbrace{\boldsymbol{\Phi}_K^\top X_i(\mathbf{t})}_{\text{PC Scores}}
 = \mathbf{P} X_i(\mathbf{t}).
$$
where \( \mathbf{P} = \bm{\Phi}_K \bm{\Phi}_K^\top \) is the projection matrix, and \( \bm{\Phi}_K \in \mathbb{R}^{p \times K} \) satisfies \( \bm{\Phi}_K^\top \bm{\Phi}_K = \mathbf{I}_K \) (i.e., the eigenvectors, by definition, are orthogonal).
In the case where both $X_i(\mathbf{t})$ and $\widehat{X}^{(K)}_{i}(\mathbf{t})$ are mean-centered, i.e.,
$$
    X_i(\mathbf{t}) = \mathbf{A} X_i(\mathbf{t})
 \quad \textbf{and} \quad 
 \widehat{X}^{(K)}_{i}(\mathbf{t}) = \mathbf{A}\widehat{X}^{(K)}_{i}(\mathbf{t})
$$
for the centering matrix \( \mathbf{A} = \mathbf{I} - \frac{1}{T} \mathbf{1} \mathbf{1}^\top \), we can expand PRESS and $\rho^2$ to get
\begin{align*}
    \text{PRESS}\bigg\{X_{i}(\mathbf{t}),  \widehat{X}^{(K)}_{i} (\mathbf{t})\bigg\} &= \bigg\{X_{i}(\mathbf{t}) -  \widehat{X}^{(K)}_{i} (\mathbf{t})\bigg\}^\top \bigg\{X_{i}(\mathbf{t}) -  \widehat{X}^{(K)}_{i} (\mathbf{t})\bigg\} \\
    &= 
    X_{i}(\mathbf{t})^\top X_{i}(\mathbf{t}) - 2 X_{i}(\mathbf{t})^\top  \widehat{X}^{(K)}_{i} (\mathbf{t}) + \widehat{X}^{(K)\top}_{i\cdot}  \widehat{X}^{(K)}_{i} (\mathbf{t}) \\
    &= X_{i}(\mathbf{t})^\top X_{i}(\mathbf{t}) - 2 X_{i}(\mathbf{t})^\top  \widehat{X}^{(K)}_{i} (\mathbf{t}) + X_{i}(\mathbf{t})^\top \boldsymbol{\Phi}_K \underbrace{\boldsymbol{\Phi}_K^\top \boldsymbol{\Phi}_K}_{= \mathbf{I}_K} \boldsymbol{\Phi}_K^\top X_{i}(\mathbf{t}) \\
    &=  X_{i}(\mathbf{t})^\top X_{i}(\mathbf{t}) - 2 X_{i}(\mathbf{t})^\top  \widehat{X}^{(K)}_{i} (\mathbf{t}) + X_{i}(\mathbf{t})^\top \underbrace{\boldsymbol{\Phi}_K \boldsymbol{\Phi}_K^\top X_{i}(\mathbf{t})}_{= \widehat{X}^{(K)}_{i} (\mathbf{t})} \\
    &= X_{i}(\mathbf{t})^\top X_{i}(\mathbf{t}) - 2 X_{i}(\mathbf{t})^\top  \widehat{X}^{(K)}_{i} (\mathbf{t}) + X_{i}(\mathbf{t})^\top  \widehat{X}^{(K)}_{i} (\mathbf{t}) \\
    &= X_{i}(\mathbf{t})^\top X_{i}(\mathbf{t}) - X_{i}(\mathbf{t})^\top  \widehat{X}^{(K)}_{i} (\mathbf{t}).
\end{align*}
Likewise, we have 
\begin{align*}
    \rho^2  \bigg\{X_{i}(\mathbf{t}),  \widehat{X}^{(K)}_{i} (\mathbf{t})\bigg\}
    =&
    \bigg\{\widehat{X}^{(K)}(\mathbf{t})X_{i}(\mathbf{t}) X_{i}(\mathbf{t})^\top \widehat{X}^{(K)}(\mathbf{t}) \bigg\}
     \bigg\{X_{i}(\mathbf{t})^\top X_{i}(\mathbf{t}) \underbrace{\widehat{X}^{(K)\top}(\mathbf{t}) \widehat{X}^{(K)}_{i} (\mathbf{t})}_{=X_{i}(\mathbf{t})^\top  \widehat{X}^{(K)}_{i} (\mathbf{t})} \bigg\}^{-1} \\
     =& 
     \bigg\{\widehat{X}^{(K)^\top}(\mathbf{t})X_{i}(\mathbf{t}) X_{i}(\mathbf{t})^\top \widehat{X}^{(K)}(\mathbf{t}) \bigg\}
     \bigg\{X_{i}(\mathbf{t})^\top X_{i}(\mathbf{t}) X_{i}(\mathbf{t})^\top  \widehat{X}^{(K)}_{i} (\mathbf{t}) \bigg\}^{-1} \\
     =&
     \widehat{X}^{(K)^\top}(\mathbf{t})X_{i}(\mathbf{t}) \underbrace{X_{i}(\mathbf{t})^\top \widehat{X}^{(K)}(\mathbf{t})
     \bigg\{X_{i}(\mathbf{t})^\top \widehat{X}^{(K)}(\mathbf{t}) \bigg\}^{-1}}_{=\mathbf{I}}
     \bigg\{X_{i}(\mathbf{t})^\top X_{i}(\mathbf{t})\bigg\}^{-1}  \\
     =& \widehat{X}^{(K)^\top}(\mathbf{t})X_{i}(\mathbf{t})  \bigg\{X_{i}(\mathbf{t})^\top X_{i}(\mathbf{t})\bigg\}^{-1} \\
     =&
     \widehat{X}^{(K)}(\mathbf{t})^\top X_{i}(\mathbf{t})  \bigg\{X_{i}(\mathbf{t})^\top X_{i}(\mathbf{t})\bigg\}^{-1} + 
     \bigg\{X_{i}(\mathbf{t})^\top X_{i}(\mathbf{t})\bigg\}  \bigg\{X_{i}(\mathbf{t})^\top X_{i}(\mathbf{t})\bigg\}^{-1} \\ &- 
     \bigg\{X_{i}(\mathbf{t})^\top X_{i}(\mathbf{t})\bigg\}  \bigg\{X_{i}(\mathbf{t})^\top X_{i}(\mathbf{t})\bigg\}^{-1} \\
     =& 
     1 - \underbrace{\bigg\{ X_{i}(\mathbf{t})^\top X_{i}(\mathbf{t}) - \widehat{X}^{(K)}(\mathbf{t})^\top X_{i}(\mathbf{t}) \bigg\}}_{= \text{PRESS}\bigg\{X_{i}(\mathbf{t}),  \widehat{X}^{(K)}_{i} (\mathbf{t})\bigg\}} \bigg\{X_{i}(\mathbf{t})^\top X_{i}(\mathbf{t})\bigg\}^{-1} \\
     =&
     1 - \text{PRESS}\bigg\{X_{i}(\mathbf{t}),  \widehat{X}^{(K)}_{i} (\mathbf{t})\bigg\} \bigg\{X_{i}(\mathbf{t})^\top X_{i}(\mathbf{t})\bigg\}^{-1}.
\end{align*}
Thus, when we use the complement of the squared correlation as our loss, we have
\begin{align*}
    \text{Loss} \left\{ X_i(\mathbf{t}) \right\}  &= 1 - \rho^2  \bigg\{X_{i}(\mathbf{t}),  \widehat{X}^{(K)}_{i} (\mathbf{t})\bigg\} \\
        &= \text{PRESS}\bigg\{X_{i}(\mathbf{t}),  \widehat{X}^{(K)}_{i} (\mathbf{t})\bigg\} \bigg\{X_{i}(\mathbf{t})^\top X_{i}(\mathbf{t})\bigg\}^{-1} \\
    &=
    \frac{\text{PRESS}\bigg\{X_{i}(\mathbf{t}),  \widehat{X}^{(K)}_{i} (\mathbf{t})\bigg\}}{\sum_{t=1}^T X_i(t)^2} = \frac{\text{PRESS}\bigg\{X_{i}(\mathbf{t}),  \widehat{X}^{(K)}_{i} (\mathbf{t})\bigg\}}{\| X_i(\mathbf{t})\|^2},
\end{align*}
which is the PRESS statistic, normalized by the squared Euclidean norm of the vector $X_i(\mathbf{t})$.
While an analogous relationship does not hold exactly when $X_i(t)$ and $\widehat{X}_i(\mathbf{t})$ are not centered, because of the non-commutativity between the centering matrix $\mathbf{A}$ and the projection matrix $\mathbf{P}$, it provides us with an intuition for $1-\rho^2$ as a measure of distance between the observed data and its predictions, that is normalized to account for the scale of the data.

\section{Additional Results of Sample Size Experiment}\label{sec:additional-results}

\begin{figure}
    \centering
    \includegraphics[width=1\linewidth]{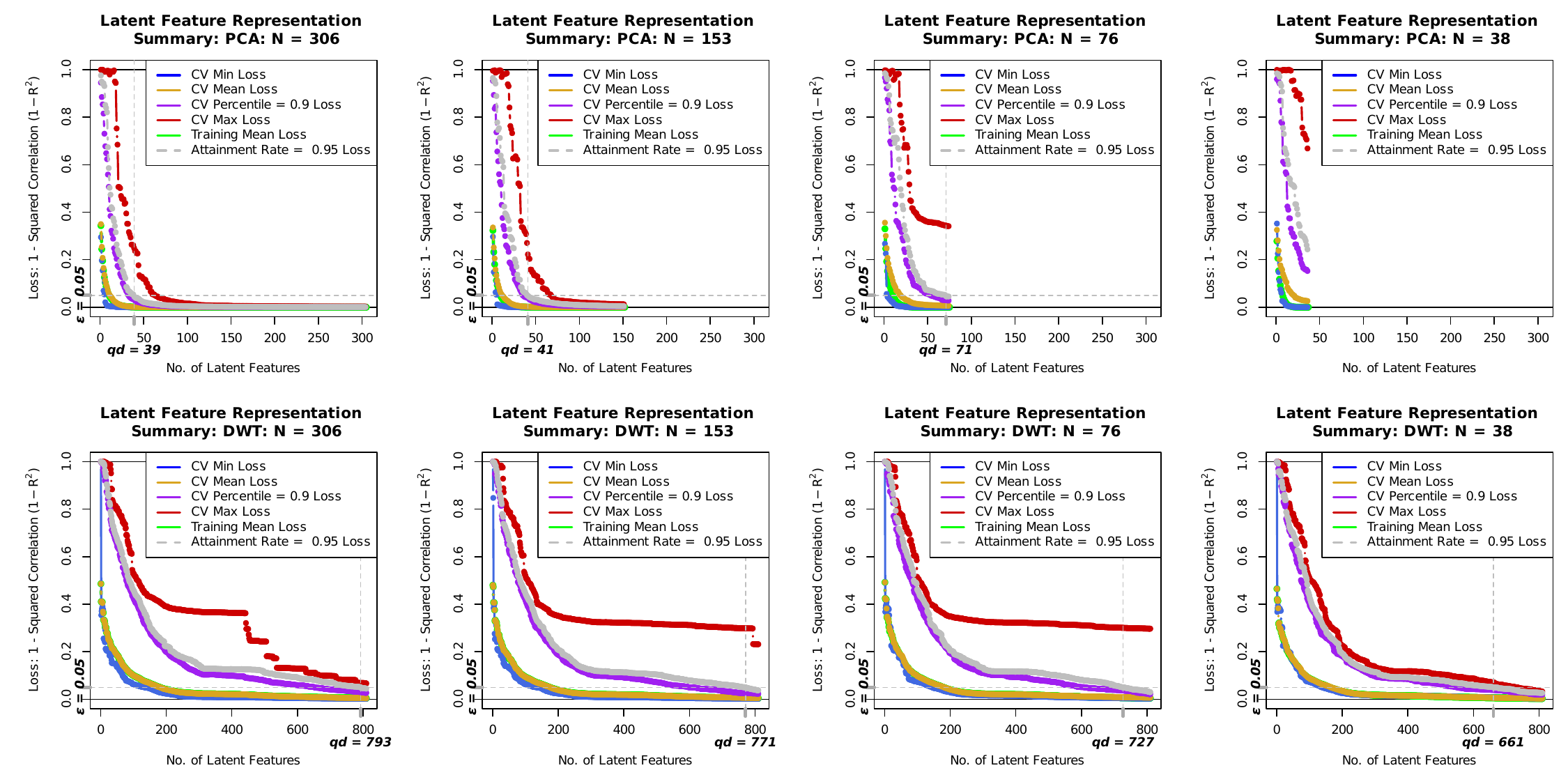}
    \caption{Results of re-running the sample size experiment presented in Figure \ref{fig:eye-sample-size-results-results-01} with a different random seed.}
    \label{fig:eye-sample-size-results-results-02}
\end{figure}

\begin{figure}
    \centering
    \includegraphics[width=1\linewidth]{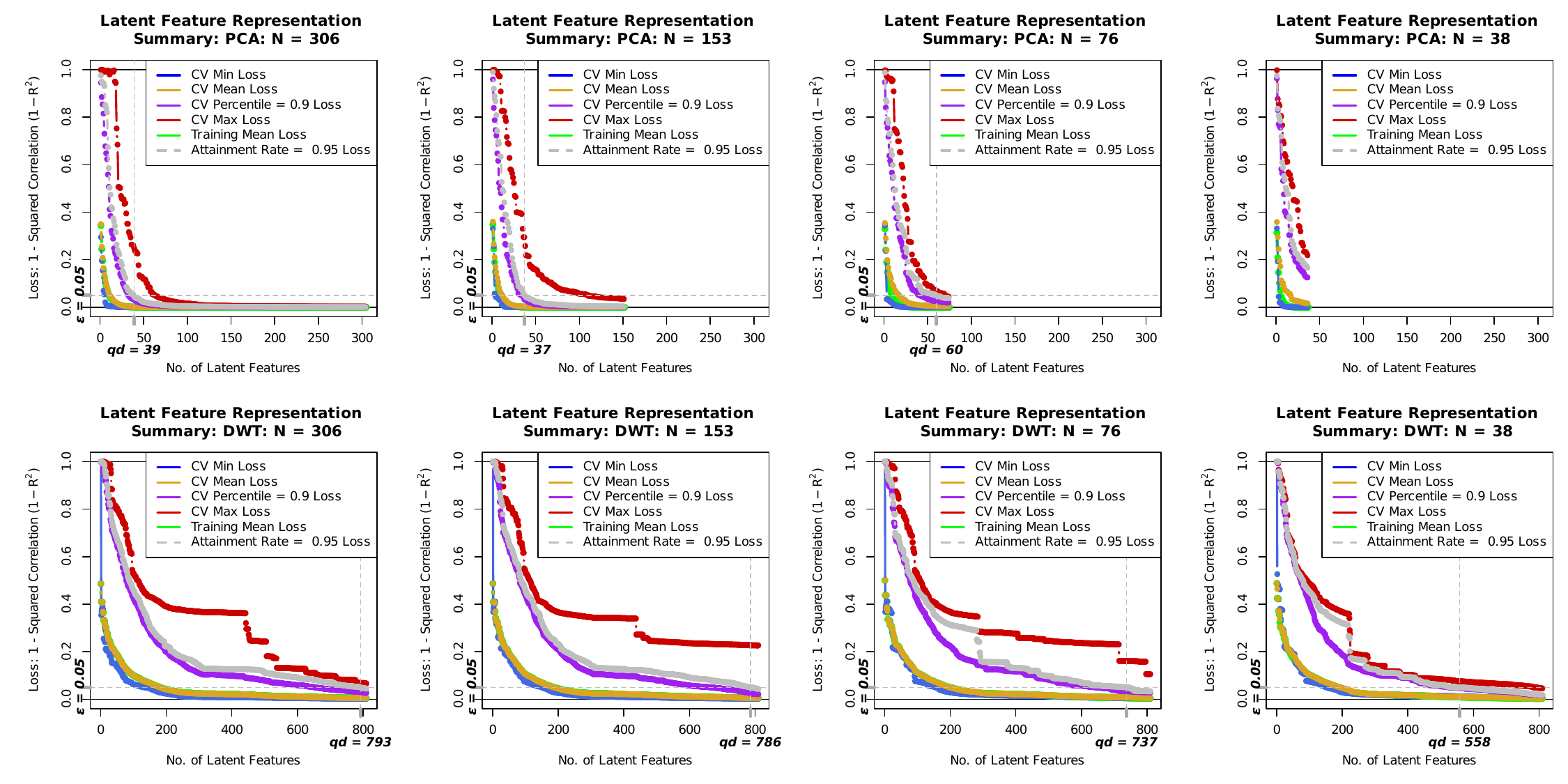}
    \caption{Results of re-running the sample size experiment presented in Figure \ref{fig:eye-sample-size-results-results-01} with a different random seed.}
    \label{fig:eye-sample-size-results-results-03}
\end{figure}

Figures \ref{fig:eye-sample-size-results-results-02} and \ref{fig:eye-sample-size-results-results-03} display the results of re-running the sample size experiment from Section \ref{sec:sample-size-experiment} with a different value of the random seed.
There are differences due to the variability induced in the sub-sampling, but the general patterns identified in Section \ref{sec:sample-size-experiment} remain the same.

\end{document}